\documentclass[10pt]{article}
\usepackage{amsmath}
\usepackage{amssymb}
\usepackage{amsmath,amsthm,amscd,amsfonts,amssymb,mathrsfs}
\usepackage[usenames]{color}
\usepackage{titlesec}

\allowdisplaybreaks
\numberwithin{equation}{section}
\titleformat*{\section}{\large\bfseries}

\newtheorem{thm}{Theorem}[section]
\newtheorem{lem}[thm]{Lemma}
\newtheorem{cor}[thm]{Corollary}
\newtheorem{prop}[thm]{Proposition}
\newtheorem{defin}[thm]{Definition}
\newtheorem{rem}[thm]{Remark}
\newtheorem{hyp}[thm]{Hypothesis}

\begin{document}
\begin{center}
{\bf Central limit theorem for the random variables associated with the IDS of the Anderson model on lattice } \\~\\
Dhriti Ranjan Dolai\\
Indian Institute of Technology Dharwad \\
Dharwad  - 580007, India.\\
Email: dhriti@iitdh.ac.in
\end{center}
{\bf Abstract:} 
We consider the existence of the integrated density of states (IDS) of the Anderson model on the Hilbert space $\ell^2(\mathbb{Z}^d)$ as analogues to the law of large numbers (LLN). In this work, we prove the analogues central limit theorem (CLT) for the collection of random variables associated with the integrated density of states for the class of test functions $C^1_P(\mathbb{R})$,  the set of all differentiable (first-order) functions on the real line whose derivative is continuous and has at most polynomial growth. Our work extends the result by Grinshpon-White (J. Spectr. Theory 12 (2022) 591-615),  where the CLT is obtained when the test functions are polynomial.
\section{Introduction}
The Anderson Model is a random Hamiltonian $H^\omega$ on $\ell^2(\mathbb{Z}^d)$ defined by
\begin{align}
\label{model}
H^\omega &=\Delta+ V^\omega,~~~\omega\in\Omega,\\
(\Delta u)(n) &=\sum_{|k-n|=1}\psi(k),~ \psi=\{\psi(n)\}_{n\in\mathbb{Z}^d}\in\ell^2(\mathbb{Z}^d),\nonumber\\
(V^\omega u)(n) &=\omega_n\psi(n) \nonumber,
\end{align}
where $\{\omega_n\}_{n\in\mathbb{Z}^d}$ are i.i.d real random variables (non-degenerate) with common distribution $\mu$.  The probability measure  $\mu$ \big(on real line\big) is known as the single site distribution (SSD). Consider the probability space $\big(\mathbb{R}^{\mathbb{Z}^d}, \mathcal{B}_{\mathbb{R}^{\mathbb{Z}^d}}, \mathbb{P} \big)$, where $\mathbb{P}=\underset{n\in\mathbb{Z}^d}{\otimes}\mu$ is constructed via the Kolmogorov theorem. We refer to this probability space as $\big(\Omega, \mathcal{B}_\Omega, \mathbb{P}\big)$ and denote $\omega=(\omega_n)_{n\in\mathbb{Z}^d}\in \Omega$.
The operator $\Delta$ is known as the discrete Laplacian, and the potential $V^\omega$ is the multiplication operator on $\ell^2(\mathbb{Z}^d)$ by the sequence $\{\omega_n\}_{n\in\mathbb{Z}^d}$.  Let $\{\delta_n\}_{n\in\mathbb{Z}^d}$ to be the standard basis for the Hilbert space $\ell^2(\mathbb{Z}^d)$.
We note that the operators $\{H^\omega\}_{\omega\in\Omega}$ are self-adjoint and have a common core domain consisting of vectors with finite support. Also, the collection $\{H^\omega\}_{\omega\in\Omega}$  is a measurable collection of random operators in the sense that for any two vectors $y, z\in\ell^2(\mathbb{Z}^d)$ the function $X_{y,z}(\omega):=\langle y,  H^\omega z\rangle: \Omega\longrightarrow \mathbb{C}$ is measurable. More about the measurability of random operators can be found in \cite{CL}.
It is well known (see \cite{pas, Kir}) that the spectrum of the random operator $H^\omega$ is a deterministic set, and it is explicitly given by $\sigma(H^\omega)=[-2d, 2d]+supp(\mu)$ a.e $\omega$.  \\~\\
Denote $\chi_{_L}$ to be  the orthogonal projection onto $\ell^2(\Lambda_L)$. Here $\Lambda_L\subset \mathbb{Z}^d$ denote the cube center at origin of side length $2L+1$, namely 
\begin{equation}
\label{box}
\Lambda_L=\big\{ n=(n_1,n_2,\cdots, n_d)  \in\mathbb{Z}^d : |n_i|\leq L\big\}~~\text{and}~~|n|=\sum_{i=1}^d |n_i|.
\end{equation}
We define the random matrix $H^\omega_L$ of size $\big|\Lambda_L\big|=(2L+1)^d$ as 
\begin{equation}
\label{restric}
H^\omega_L=\chi_{_L}H^\omega\chi_{_L}. 
\end{equation}
Now, we will describe the existence of the integrated density of states (IDS), and its proof is given in \cite[Theorem 3.15]{AMWS} (see also \cite{pas1}). Let the self-adjoint operators $H^\omega$ and $H^\omega_L$ as defined above then for any $f\in C_0(\mathbb{R})$, the set of all continuous functions on $\mathbb{R}$ which decay to zero at infinity, the limit
\begin{equation}
\label{ids}
\lim_{L\to\infty}\frac{1}{|\Lambda_L|}\sum_{n\in\Lambda_L}\langle \delta_n, f(H^\omega_L) \delta_n\rangle
=\mathbb{E}\big( \langle \delta_0, f(H^\omega) \delta_0\rangle \big)~a.e~\omega,
\end{equation}
is well known to exist. The probability measure, $\nu(\cdot)=\mathbb{E}\big( \langle \delta_0, E_{_{H^\omega}}(\cdot) \delta_0\rangle \big)$ is known as the density of states measures (DOSm). Its distribution function, $N(x)=\nu(-\infty, x],~x\in\mathbb{R}$, is known as the integrated density of states (IDS). 
Here,  $E_{_{T}}(\cdot)$ denotes the spectral measure of a self-adjoint operator $T$ defined on a Hilbert space $\mathcal{H}$ and for any two vectors 
$y,  z\in\mathcal{H}$,  the quantity $\langle y, f(T) z\rangle$ is defined through the functional calculus. Details about spectral measure and functional calculus can be found in \cite{RS}.\\
In the appendix Lemma \ref{poly-gr-weak}, we show that under the moment condition (\ref{mnts}) (see below) on the $\mu$ (SSD), the above convergence (\ref{ids}) will also hold for a larger class of test function namely,  $f\in C_P(\mathbb{R})$,  $f$ is continuous on $\mathbb{R}$ such that $|f(x)|\leq P(x)~\forall~x\in\mathbb{R}$, for some polynomial $P$.\\~\\
We note that for $f\in C_0(\mathbb{R})$, generally the collection of random variables $\{\langle \delta_n, f(H^\omega_L) \delta_n\rangle \}_{n\in\Lambda_L}$ is not independent, also it depends on $L$. But we regard the limit (\ref{ids}) as analogous to the strong law of large numbers 
(SLLN) for this collection.  Here, we are interested in finding out whether the analogous central limit theorem (CLT) for the collection of random variables
$\{\langle \delta_n, f(H^\omega_L) \delta_n\rangle \}_{n\in\Lambda_L}$ will hold or not.  It is expected that for a large class of test functions
$f$, the convergence (in the distribution sense) of random variables
\begin{equation}
\label{clt}
\frac{1}{|\Lambda_L|^{\frac{1}{2}}}\sum_{n\in\Lambda_L}\bigg(\langle \delta_n, f(H^\omega_L) \delta_n\rangle- 
\mathbb{E}\big (\langle \delta_n, f(H^\omega_L) \delta_n\rangle \big) \bigg)
\xrightarrow[L\to\infty]{\substack{in\\ distribution}} \mathcal{N}(0,\sigma^2_{f}),
\end{equation}
 will happen for some $\sigma^2_f>0$. Here $\mathcal{N}(0,\sigma^2_{f})$ denote the normal distribution with 
 zero mean and variance $\sigma^2_f$ (it is also known as the limiting variance).\\~\\
Very recently,  the central limit theorem (\ref{clt}) was obtained by Grinshpon-White \cite{GYWJ} for the model defined on $\ell^2(\mathbb{Z}^d),~d\ge 1$ when the test function is a polynomial. The authors have provided the explicit conditions on SSD for which the limiting variance $\sigma^2_P$ will be strictly positive for all polynomials and also characterised all those polynomials and SSDs for which the limiting variance $\sigma^2_P$ will be zero.\\~\\
In this work, we generalize (in terms of test functions) the result of Grinshpon-White \cite{GYWJ} and prove the central limit theorem (\ref{clt}) on $\ell^2(\mathbb{Z}^d), d\ge 1$, when the test function $f$ is in $C_P^1(\mathbb{R})$ (see Definition \ref{def}) and for proving that we required some conditions on the growth of the moments of the SSD. The CLT for the class of test functions $C_P^1(\mathbb{R})$ is valid even if single site distribution (SSD) has a singular component. Namely, the CLT (\ref{m-clt-dif-f}) (below) is true even if the Bernoulli distribution acts as the SSD. This work does not assume any localization (or spectral) properties to prove our result.\\~\\
Apart from \cite{GYWJ}, all the other earlier results are valid for the one-dimensional model, and they can be found in \cite{Rez}, \cite{KP}, \cite{pas-sh}. Similar questions have been studied in random matrix theory with great detail; we refer to \cite{fos},  \cite{PLSM} and \cite{AGZ} and reference therein as general references.\\~\\
In \cite{Rez}, Reznikova considered random Schr\"{o}dinger operator on $\ell^2(\mathbb{N}\cup \{0\})$ with compactly supported absolutely continuous single site distribution (SSD). It was proved that the random  process 
$$ N^*_L(E)=\frac{N_L(E)-LN(E)}{\sqrt{L}}$$
converges to a Gaussian process as $L$ gets large in the sense of convergence of finite-dimensional distributions. Here $N_L(E)$ count the number of eigenvalues of the restriction (of the full operator) to $\ell^2\{0,1,2,\cdots, L\}$ which are below $E$ and $N(E)$ be the value of the IDS at the point $E$. The result described above will give the equivalence of the convergence (\ref{clt}) when $f(x)=\chi_{(-\infty, E]}(x)$,  the characteristic function of the interval $(-\infty, E],~E\in\mathbb{R}$.
We also refer to \cite{R, R2} for the one-dimensional continuous model.\\
Kirsch-Pastur \cite{KP} considered the model on $\ell^2(\mathbb{Z})$ with compactly supported SSD and proved the central limit theorem \big(as described in (\ref{clt})\big) for the function $f(x)=(x-E)^{-1}$ for some $E$ satisfy $\text{dist}(E, \sigma(H^\omega))>0$.  Let $\hat{f}(t)$, the Fourier transform of the function $f$, which has sufficient decay at infinity \big(i.e. $f$ has sufficiently higher order of smoothness\big), then for this test function $f$ the CLT (\ref{clt}) was obtained by Pastur-Shcherbina \cite{pas-sh} on $\ell^2(\mathbb{Z})$ and they also provided the criterion for the positivity of the limiting variance $\sigma^2_f$. \\
For the Schr\"{o}dinger operator on $\ell^2({\mathbb{N}})$ with decaying random potential (non-stationary), the convergence (\ref{clt}) was shown by Breuer et al. \cite{BGW}, when the test function $f$ is a polynomial. By Mashiko et al.  \cite{MMMN} for real analytic test function.\\~\\
Let us define the class of functions $C_P^1(\mathbb{R})$ formally before making the hypothesis to prove the central limit theorem for this class. 
\begin{defin}
\label{def}
We say $f\in C_P^1(\mathbb{R})$ if $f$ is a real-valued differentiable function on the real line $\mathbb{R}$ and its derivative (first-order) is continuous on $\mathbb{R}$ such that $\big|f'(x)\big|\leq P(x)~\forall~x\in\mathbb{R}$, for some polynomial $P(x)$.
\end{defin}
\begin{rem}
Let $f$ be a differentiable real-valued function on the real line such that $f'$ is continuous on $\mathbb{R}$ and also $|f'(x)|\to 0$ as $|x| \to \infty,$ then $f$ is in the class $C_P^1(\mathbb{R})$ as defined in the above.
\end{rem}
\noindent We set a few notations to state the CLT for the class of test functions $ C^1_P(\mathbb{R})$.  For a real-valued function (Borel-measurable) $f$ the random variable $X_{f,L}$ is defined by
\begin{align}
\label{sm-rv-diff}
 X_{f, L}(\omega):&=\frac{1}{|\Lambda_L|^\frac{1}{2}}
\sum_{n\in\Lambda_L}
  \bigg(\big\langle \delta_n, f(H^\omega_L)\delta_n\big\rangle-
\mathbb{E}\big(\langle \delta_n, f(H^\omega_L)\delta_n\rangle  \big)\bigg)\nonumber\\
&=\frac{1}{|\Lambda_L|^\frac{1}{2}}\bigg( Tr\big(f\big( H^\omega_L\big)\big)-
\mathbb{E}\bigg(  Tr\big(f\big( H^\omega_L\big)\big) \bigg) \bigg).
\end{align}
Also we define $\sigma^2_f$, the limiting variance (as $L$ gets large) of the sequence of random variables $\big\{ X_{f,L} \big\}_L$ as
\begin{equation}
\label{df-lm-vr}
\sigma^2_f:=\displaystyle\limsup_{L\to\infty}\mathbb{E}\big( \big| X_{f,L} \big|^2 \big)
\end{equation}
In Proposition \ref{fin-diff}, we showed the finiteness of $\sigma_f^2$ for every $f\in C_P^1(\mathbb{R})$ and to do so, we required the moment determinacy of the modified density of states measure (modified DOSm) $\bar{\nu}$ as defined by (\ref{full-ldosm}) so that we can use that fact that polynomials are dense in $L^2(\bar{\nu})$.  Also, the denseness of polynomials in $L^2(\bar{\nu})$ is helpful to show the positivity of $\sigma^2_f$ as well, see Corollary (\ref{pol-dens}).  To prove that the $\bar{\nu}$ (modified DOSm) is determined by its moments, we assume some growth conditions on the moments (absolute) of the $\mu$ (SSD).
\begin{hyp}
\label{hp2}
The single site distribution (SSD) $\mu$ has all the moments, and it satisfies the conditions
\begin{equation}
\label{mnts}
\int |x|^k d\mu(x) \leq C a^k k^k~~\forall~k\in\mathbb{N},~\text{for some}~C, a\geq 1.
\end{equation}
\end{hyp}
\begin{rem}
\label{exple}
The normal distribution and any compactly supported probability measure on the real line $\mathbb{R}$ will quickly satisfy the above conditions (\ref{mnts}) on its absolute moments.
\end{rem}
\begin{rem}
We took the explicit growth rate (\ref{mnts}) of the moments of $\mu$ (SSD) to show the moment determinacy of the $\bar{\nu}$ \big(modified DOSm, see (\ref{full-ldosm})\big) using the expression (\ref{spt}), (\ref{mono}) and (\ref{mono-fn}).  Suppose for some other condition (less restrictive) on $\mu$ (SSD), the $\bar{\nu}$ (modified DOSm) is determined by its moments; then, in that case, under the very same condition, the Theorem \ref{clt-for-dif} (below) will also be true.
\end{rem}
\noindent Now, we are ready to state the main result of our work:
\begin{thm}
\label{clt-for-dif}
Let $f\in C_P^1(\mathbb{R})$ and consider $H^\omega$, $H^\omega_L$ as define in (\ref{model}), (\ref{restric}) then under the Hypothesis \ref{hp2} we have
\begin{equation}
\label{m-clt-dif-f}
\frac{1}{|\Lambda_L|^{\frac{1}{2}}}\sum_{n\in\Lambda_L}\bigg(\langle \delta_n, f(H^\omega_L) \delta_n\rangle- 
\mathbb{E}\big (\langle \delta_n, f(H^\omega_L) \delta_n\rangle \big) \bigg)
\xrightarrow[L\to\infty]{\substack{in\\ distribution}} \mathcal{N}(0,\sigma^2_{f}),
\end{equation}
here $\mathcal{N}(0,\sigma^2_{f})$ is the normal distribution with zero mean and variance $\sigma^2_f$, as given in (\ref{df-lm-vr}).  The limiting variance $\sigma^2_f$ is always finite for any $f\in C^1_P(\mathbb{R})$.  We also show that the variance $\sigma^2_f$ is strictly positive for those $f\in C^1_P(\mathbb{R})$ which are strictly monotone functions on an open interval $I$ (non-random) such that $\sigma(H^\omega)\subseteq I~a.e~\omega$,  here $\sigma(H^\omega)$ is the spectrum of $H^\omega$, it is deterministic a.e $\omega$.
\end{thm}
\begin{rem}
When the single site distribution (SSD) $\mu$ has unbounded support, the open interval $I$ (non-random) will also be unbounded.
\end{rem}
\begin{rem}
We observe that if the test function $f\in C^1_P(\mathbb{R})$ is strictly monotone on the open interval $I$ (non-random), which contains $\sigma(H^\omega)$, then to prove the limiting variance $\sigma^2_f$ is positive (strictly) we do not need any restriction on the cardinality of the support of $\mu$ (SSD) (on the other hand, it is required when the test function is a polynomial).  In particular, for strictly monotone $\big(\text{on}~I\supseteq \sigma(H^\omega)\big)$ test function $f\in C^1_P(\mathbb{R})$ the limiting variance $\sigma^2_f>0$ even if the single site distribution (SSD) $\mu$ is Bernoulli.
\end{rem}
\begin{rem}
In the above if $\sigma^2_f=0$ for some function $f$ in $C_P^1(\mathbb{R})$ then limit for the convergence (in distribution) in
(\ref{m-clt-dif-f}) will be zero, instead of $\mathcal{N}(0,\sigma^2_{f})$. 
\end{rem}
\noindent Since the central limit theorem (CLT) is valid for all real polynomials as a test function (\cite{GYWJ}),  then the natural expectation is that the same  (\ref{clt}) should also hold for each $f\in C_c(\mathbb{R})$, set of all compactly supported continuous functions on the real line $\mathbb{R}$.  
Let $\{P_n \}_{n=1}^\infty$ be a sequence of polynomials which converges uniformly to a compactly supported continuous function $f$. One way to approximate the CLT of $f$ by CLTs of $\{P_n\}_n$ is through the Theorem \ref{app-con-thm} below. To apply the Theorem \ref{app-con-thm} estimation of the variance of the random variable $X_{(f-P_n),  L}$ (given in (\ref{sm-rv-diff}) below) is very crucial. In fact, we need to show the limit 
\begin{equation}
\label{double-limi}
\lim_{n\to\infty}\limsup_{L\to\infty}\mathbb{E}\bigg( \bigg|X_{(f-P_n), L}\bigg|^2 \bigg)=0.
\end{equation}
The above limit is not a straightforward application of the Wierstrass approximation theorem.  But for a differentiable (first-order) function whose derivative grows at most in polynomial order, the limit (\ref{double-limi}) can be achieved with the help of martingale-difference sequence and the derivative formula for trace as in Lemma \ref{der-tr} (in the appendix).\\~\\
Our method is based on the moments analysis of the collection of random variables $\big\{\langle \delta_n, f(H^\omega_{\Lambda})\delta_n\rangle\big\}_{n\in\Lambda}$, $\Lambda\subseteq\mathbb{Z}^d$.   For $f\in C^1_P(\mathbb{R})$ using the fundamental theorem of calculus, we can always find a real polynomial $\tilde{P}$ such that $|f(x)|\leq |\tilde{P}(x)|$, $\forall~x\in\mathbb{R}$ therefore using the Hypothesis \ref{hp2} and Remark \ref{poly-op} it is easy to claim that all the moments of the random variable $\langle \delta_n, f(H^\omega_{\Lambda})\delta_n\rangle$ exists for each $n\in\mathbb{Z}^d$.\\
In \cite{GYWJ}, the authors have used the path counting method to compute the trace of powers of $H^\omega_L$ for proving the CLT (\ref{clt}) when the test function is a polynomial.\\
Our proof of the Theorem \ref{clt-for-dif} relies the moment's analysis of the random variable $\langle \delta_n, f'(H^\omega_{\Lambda})\delta_n\rangle$,  $f\in C^1_P(\mathbb{R})$.  We also required an explicit lower bound (involving the expression of $P$) of the $\sigma^2_P$ (limiting variance when the test function is a polynomial $P$) to prove the positivity of $\sigma^2_f$; therefore, to make our work more understandable, we are going to describe the result of \cite{GYWJ} explicitly.\\~\\
\noindent For the central limit theorem (CLT) of polynomials as a test function, we will assume the existence of all the moments of the single site distribution (SSD). 
\begin{hyp}
\label{hp}
 The single site distribution (SSD) $\mu$ has all the moments.
\end{hyp}
\noindent 
We refer to \cite{GYWJ} for the proof of the following theorem.
\begin{thm}
\label{main}
Let $P$ is a real polynomial of degree $p\ge 1$ and $H^\omega$, $H^\omega_L$ as defined in (\ref{model}),  (\ref{restric}), 
then under the Hypothesis \ref{hp} we have
\begin{equation}
\label{m-clt}
\frac{1}{|\Lambda_L|^{\frac{1}{2}}}\sum_{n\in\Lambda_L}\bigg(\langle \delta_n, P(H^\omega_L) \delta_n\rangle-
\mathbb{E}\big (\langle \delta_n, P(H^\omega_L) \delta_n\rangle \big) \bigg)
\xrightarrow[L\to\infty]{\substack{in\\ distribution}} \mathcal{N}(0,\sigma^2_{P}).
\end{equation}
In the above $\mathcal{N}(0,\sigma^2_{P})$ denote the normal distribution with zero mean and variance $\sigma^2_P$.  The limiting variance $\sigma^2_P$ is always finite and non-negative.
\end{thm}
\begin{rem}
More details about the limiting variance $\sigma^2_P$ can be found in \cite{GYWJ}.
\end{rem}
\noindent Now, before going to the section on the proof of our result Theorem \ref{clt-for-dif}, we set a few notations that will be used in the rest of the paper.\\
\noindent Let $\Lambda^{int}_{L,p},~L>p$ be an interior of the cube $\Lambda_L$ defined by
\begin{equation}
\label{inter}
\Lambda^{int}_{L,p}:=\big\{n=(n_1,n_2,\cdots, n_d)\in\Lambda_L: |n_i|<L-p  \big \},~~L\in\mathbb{N}.
\end{equation}
Since $P$ is a polynomial (real) of degree $p$ then, for each $n\in \Lambda^{int}_{L,p}$ using the definitions (\ref{model}) and (\ref{restric}) we always have $P(H^\omega_L)\delta_n=P(H^\omega)\delta_n$.\\
Lets denote $\Lambda_\infty=\mathbb{Z}^d$ and $H^\omega=H^\omega_{\Lambda_\infty}$.
Now for any $u\in\mathbb{R}$ and $B\subseteq \Lambda_L\subseteq \mathbb{Z}^d$,  for $L\in \mathbb{N}\cup\{\infty\}$,  we defined the modified potential $V^\omega_L\big|_{(\omega_j\to u\omega_j)_{j\in B}}$ on $\ell^2(\Lambda_L)$ as
\begin{equation}
\label{chg-ssd-B}
\bigg(V^\omega_L\big|_{(\omega_j\to u\omega_j)_{j\in B}} \psi\bigg)(n)=\left\{
 \begin{array}{rl}
  (u\omega_n) \psi(n) & \text{if } n\in B \\
   \omega_n \psi(n) & \text{if } n\in \Lambda_L\setminus B.
   \end{array} \right.
\end{equation}
In the above $\psi\in\ell^2(\Lambda_L)$. The self-adjoint operator $H^\omega_L\big|_{(\omega_j\to u\omega_j)_{j\in B}}$ on $\ell^2(\Lambda_L)$ is defined by
\begin{equation}
\label{md-chn-ssd}
H^\omega_L\big|_{(\omega_j\to u\omega_j)_{j\in B}}=\chi_L\Delta\chi_L+V^\omega_L\big|_{(\omega_j\to u\omega_j)_{j\in B}}.
\end{equation}
  For simplicity in the writing, we denote $\big\langle \delta_n,  E_{H^\omega_L}(\cdot)\delta_n\big\rangle\big|_{(\omega_j\to u\omega_j)_{j\in B}}$ to be the spectral measure of the operator $H^\omega_L\big|_{(\omega_j\to u\omega_j)_{j\in B}}$ at the vector $\delta_n$.  
In the case of singleton $B=\{j\}$ we write $H^\omega_L\big|_{(\omega_j\to u\omega_j)_{j\in B}}=H^\omega_L\big|_{(\omega_j\to u\omega_j)}$.
 Let $g$ be a real-valued Borel-measurable function on $\mathbb{R}$.  Now, using the spectral theorem of the operator $H^\omega_L\big|_{(\omega_j\to u\omega_j)_{j\in B}}$ we define
  \begin{equation}
  \label{sp-thm-chn-ssd}
\big\langle \delta_n,  g(H^\omega_L)\delta_n\big\rangle\big|_{(\omega_j\to u\omega_j)_{j\in B}}
=\int_{\mathbb{R}}g(x)~d\big\langle \delta_n,  E_{H^\omega_L}(x)\delta_n\big\rangle\big|_{(\omega_j\to u\omega_j)_{j\in B}}.
\end{equation}
\section{Proof of the CLT when test functions are in $C^1_P(\mathbb{R})$}
\noindent Our object here is to prove the Theorem \ref{clt-for-dif} i.e.  CLT (\ref{clt}) for a larger class of test functions, namely $C^1_{P}(\mathbb{R})$, is described in the Definition \ref{def}.  We will obtain the CLT for $f\in C^1_{P}(\mathbb{R})$ as a limit of the CLTs of polynomials, and for that, the following theorem is essential.  
\begin{thm}
\label{app-con-thm}
Let $\{Y_n\}_{n=1}^\infty$ and $\{Z_{n,k}\}_{n,k=1}^\infty$ are real-valued random variables.  Assume that
\begin{center}
\begin{enumerate}
\item [(a)] $Z_{n,k}\xrightarrow[n\to\infty]{\text{in distribution}}Z_k$,~ for each fix $ k$.\\
\item [(b)] $Z_k\xrightarrow[k\to\infty]{\text{in distribution}}Z$.\\
\item [(c)] For each $\delta>0$,  $\displaystyle \lim_{k\to\infty}\limsup_{n\to\infty}\mathbb{P}\bigg( \big|Z_{n,k}-Y_n  \big|\ge \delta \bigg)=0$.
\end{enumerate}
\end{center}
Then $Y_n\xrightarrow[n\to\infty]{\text{in distribution}}Z$. 
\end{thm}
\noindent The proof of the above result can be found in \cite[Theorem 25.5]{PB}.\\~\\
First, we will obtain a uniform bound (independent of $L$) for the variances of the random variables
$\{X_{f,L}\}_L$, given in (\ref{sm-rv-diff}).   Before doing so, lets introduce two finite measures $\bar{\nu}_L(\cdot)$ and $\bar{\nu}(\cdot)$ with the help of the notations (\ref{md-chn-ssd}) and
(\ref{sp-thm-chn-ssd}) \big(when $B$ is singleton\big) as
\begin{equation}
\label{ldosm}
\bar{\nu}_L\big(\cdot\big)=\frac{1}{|\Lambda_L|}\sum_{n\in\Lambda_L}\int_0^1\big[\mathbb{E}\big(\omega^2_n~\big\langle \delta_n, E_{H^\omega_L} \big(\cdot \big) \delta_n  \big\rangle\big|_{(\omega_n\to u\omega_n)}\big)\big]du.
\end{equation}
For the full operator $H^\omega$,  we also define
\begin{equation}
\label{full-ldosm}
\bar{\nu}\big(\cdot\big)=\int_0^1\big[\mathbb{E}\big(\omega^2_0~\big\langle \delta_0, E_{H^\omega} \big(\cdot \big) \delta_0  \big\rangle\big|_{(\omega_0\to u\omega_0)}\big)\big]du.
\end{equation}
Both the above measures $\bar{\nu}_L$ and $\bar{\nu}$ are finite, namely $\bar{\nu}_L(\mathbb{R})=\bar{\nu}(\mathbb{R})=\mathbb{E}(\omega_0^2)$.  Also,  in the Lemma \ref{weak-con} (appendix) we show that the sequence of measure $\big\{\bar{\nu}_L \big\}_L$ converges weakly to $\bar{\nu}$.\\
Let the single site distribution $\mu$ has all the moments as in (\ref{mnts}), then it is given in Lemma \ref{weak-con} (in Appendix) that $f'\in L^2(\bar{\nu})\cap L^2(\bar{\nu}_L)$ for $f\in C_P^1(\mathbb{R})$.  Here $\bar{\nu}=\bar{\nu}_p$ and $\bar{\nu}_{L}=\bar{\nu}_{L,p}$ for $p=1$, see (\ref{re-ldosm}) and (\ref{fl-re-ldosm}) in appendix.
\begin{prop}
\label{fin-diff} 
Under the Hypothesis \ref{hp2} for any $f\in C_{P}^1({\mathbb{R}})$ we always have 
$$0\leq \sigma^2_f\leq 8\| f' \|^2_{L^2(\bar{\nu})}<\infty,~~~\sigma^2_f=\limsup_{L\to\infty}\mathbb{E}\big(\big|X_{f,L}\big|^2 \big),  $$
where $X_{f, L}(\omega):=\displaystyle \frac{1}{|\Lambda_L|^\frac{1}{2}}\sum_{n\in\Lambda_L}\bigg(\big\langle \delta_n, f(H^\omega_L)\delta_n\big\rangle-
\mathbb{E}\big(\big\langle \delta_n, f(H^\omega_L)\delta_n\big\rangle  \big)\bigg)$. 
\end{prop}
\begin{proof}
Here, we are also going to use the martingale difference techniques.  Let $\{n_k\}_{k=1}^{(2L+1)^d}$ ($n_k<n_{k+1}$) be an enumeration of all the elements of the finite box $\Lambda_L\subset \mathbb{Z}^d$,  here $\big|\Lambda_L \big|=(2L+1)^d$,  see (\ref{box}).  Now we define a filtration $\{\mathcal{G}_k \}_{k=1}^{(2L+1)^d}$ of $\sigma$-algebras 
\begin{equation}
\label{g-algb}
\mathcal{G}_k=\sigma\big(\omega_n: 1\leq n\leq n_k \big)~~\text{and we have}~~\mathcal{G}_k\subset \mathcal{G}_{k+1}.
\end{equation}
Lets define $\Psi_{f,L}(\omega)$, the trace of $f(H^\omega_L)$ as
\begin{equation}
\label{tr-fun}
\Psi_{f,L}(\omega):=\sum_{n\in\Lambda_L}\big\langle \delta_n, f(H^\omega_L)\delta_n\big\rangle=Tr\big(f(H^\omega_L)  \big)
\end{equation}
Now it is easy to check that $\big\{\mathbb{E}\big(\Psi_{f,L}\big| \mathcal{G}_k  \big) \big\}_{k=1}^{(2L+1)^d} $ is a martingale (Doob) w.r.t the filtration $\{\mathcal{G}_k \}_{k=1}^{(2L+1)^d}$.  Since $\Psi_{f,L}(\omega)$ is function of $\{\omega_n\}_{n\in\Lambda_L}$ only, so we get $\mathbb{E}\big(\Psi_{f,L}\big| \mathcal{G}_{(2L+1)^d}  \big)=\Psi_{f,L}(\omega)$ and denote $\mathbb{E}\big(\Psi_{f,L}\big| \mathcal{G}_0  \big)=\mathbb{E}\big(\Psi_{f,L}(\omega)  \big)$,  here $\mathcal{G}_0$ denote the trivial $\sigma$-algebra (consists of empty and total space).  Now we write the difference between $\Psi_{f,L}$ and its expectation as
\begin{equation}
\label{fn-mrt-dif}
\Psi_{f,L}(\omega)-\mathbb{E}\big(\Psi_{f,L}(\omega)  \big)=\sum_{k=1}^{(2L+1)^d}\bigg(\mathbb{E}\big(\Psi_{f,L}\big| \mathcal{G}_k  \big)-\mathbb{E}\big(\Psi_{f,L}\big| \mathcal{G}_{k-1}  \big)  \bigg).
\end{equation}
Since the covariance between any two distinct elements from a martingale (Doob) difference sequence is always zero,  see Proposition \ref{or-db-dif-mart}, in the appendix, then the variance of the random variable $X_{f,L}(\omega)$ can be written as
\begin{align}
\label{fun-mrt-dis}
\mathbb{E}\big(\big| X_{f,L} \big| ^2 \big)&=\frac{1}{|\Lambda_L|}\mathbb{E}\bigg( \Psi_{f,L}(\omega)-\mathbb{E}\big(\Psi_{f,L}(\omega)  \big) \bigg)^2\nonumber\\
&=\frac{1}{|\Lambda_L|}\sum_{k=1}^{(2L+1)^d}\mathbb{E}\bigg(\mathbb{E}\big(\Psi_{f,L}\big| \mathcal{G}_k  \big)-\mathbb{E}\big(\Psi_{f,L}\big| \mathcal{G}_{k-1}  \big)  \bigg)^2.
\end{align}
Now, we will estimate each term inside the r.h.s of the above sum.
\begin{align}
\label{fun-dif-est}
&\mathbb{E}\big(\Psi_{f,L}\big| \mathcal{G}_k  \big)-\mathbb{E}\big(\Psi_{f,L}\big| \mathcal{G}_{k-1}  \big),~~
\text{here}~\Psi_{f,L}:=\Psi_{f,L}(\omega),~\omega=(\omega_n)_{n\in\mathbb{Z}^d}\nonumber\\
&\qquad\qquad\qquad =\mathbb{E}\big(\Psi_{f,L}\big| \mathcal{G}_k  \big)-
\mathbb{E}\big(\Psi_{f,L}\big|_{(\omega_{n_k}=0)}\big| \mathcal{G}_{k}  \big)
\nonumber\\
&\qquad\qquad \qquad\qquad -\bigg(\mathbb{E}\big(\Psi_{f,L}\big| \mathcal{G}_{k-1}  \big)-
\mathbb{E}\big(\Psi_{f,L}\big|_{(\omega_{n_k}=0)}\big| \mathcal{G}_{k} \big)\bigg)\nonumber\\
&\qquad\qquad\qquad =\mathbb{E}\big(\Psi_{f,L}\big| \mathcal{G}_k  \big)-
\mathbb{E}\big(\Psi_{f,L}\big|_{(\omega_{n_k}=0)}\big| \mathcal{G}_{k}  \big)
\nonumber\\
&\qquad\qquad \qquad\qquad -\bigg(\mathbb{E}\big(\Psi_{f,L}\big| \mathcal{G}_{k-1}  \big)-
\mathbb{E}\big(\Psi_{f,L}\big|_{(\omega_{n_k}=0)}\big| \mathcal{G}_{k-1} \big)\bigg)\nonumber\\
&\qquad\qquad\qquad=\mathbb{E}\bigg(\int_0^1\frac{d}{du}\big(\Psi_{f,L}\big|_{(\omega_{n_k}\to u\omega_{n_k})}\big| \mathcal{G}_k  \big)du\bigg)\nonumber\\
&\qquad\qquad \qquad\qquad -\mathbb{E}\bigg(\int_0^1\frac{d}{du}\big(\Psi_{f,L}\big|_{(\omega_{n_k}\to u\omega_{n_k})}\big| \mathcal{G}_{k-1}  \big)du\bigg)\nonumber\\
&\qquad\qquad\qquad=\int_0^1\bigg[\mathbb{E}\bigg(\omega_{n_k}\big\langle \delta_{n_k},  f'(H^\omega_L)\delta_{n_k}\big\rangle\big|_{(\omega_{n_k}\to u\omega_{n_k})}\big|\mathcal{G}_k \bigg)\nonumber\\
&\qquad\qquad \qquad\qquad-\mathbb{E}\bigg(\omega_{n_k}\big\langle \delta_{n_k},  f'(H^\omega_L)\delta_{n_k}\big\rangle\big|_{(\omega_{n_k}\to u\omega_{n_k})}\big|\mathcal{G}_{k-1} \bigg)\bigg]du.
\end{align}
In the last line of the above, we have used the formula for the derivative of the trace of $f'(H^\omega_L)$ as in (\ref{tr-fr}), in the appendix. Given the Remark \ref{rs-fr-chg-int} (below), we have applied Fubini's theorem to change the order of integration above. Now, using Jensen's inequality for conditional expectation and the inequality $(a+b)^2\leq 4(a^2+b^2)$ (real $a$, $b$), we estimate the above as
\begin{align*}
&\bigg(\mathbb{E}\big(\Psi_{f,L}\big| \mathcal{G}_k  \big)-\mathbb{E}\big(\Psi_{f,L}\big| \mathcal{G}_{k-1}  \big)  \bigg)^2\\
&\qquad \qquad \qquad\leq 4\bigg[\bigg( \int_0^1\bigg[\mathbb{E}\bigg(\omega_{n_k}\big\langle \delta_{n_k},  f'(H^\omega_L)\delta_{n_k}\big\rangle\big|_{(\omega_{n_k}\to u\omega_{n_k})}\big|\mathcal{G}_k \bigg)\bigg] du\bigg)^2  \\
&\qquad\qquad\qquad\qquad+\bigg( \int_0^1\bigg[\mathbb{E}\bigg(\omega_{n_k}\big\langle \delta_{n_k},  f'(H^\omega_L)\delta_{n_k}\big\rangle\big|_{(\omega_{n_k}\to u\omega_{n_k})}\big|\mathcal{G}_{k-1} \bigg)\bigg]du \bigg)^2 \bigg]\\
&\qquad \qquad \qquad\leq 4\bigg[\bigg( \int_0^1\bigg[\mathbb{E}\bigg(\omega^2_{n_k}\big(\big\langle \delta_{n_k},  f'(H^\omega_L)\delta_{n_k}\big\rangle\big)^2\big|_{(\omega_{n_k}\to u\omega_{n_k})}\big|\mathcal{G}_k \bigg)\bigg] du\bigg)   \\
&\qquad\qquad\qquad\qquad+\bigg( \int_0^1\bigg[\mathbb{E}\bigg(\omega^2_{n_k}\big(\big\langle \delta_{n_k},  f'(H^\omega_L)\delta_{n_k}\big\rangle\big)^2\big|_{(\omega_{n_k}\to u\omega_{n_k})}\big|\mathcal{G}_{k-1} \bigg)\bigg]du \bigg) \bigg]
\end{align*}
Now, using the total law of expectation above, we get
\begin{align*}
\mathbb{E}\bigg(\mathbb{E}\big(\Psi_{f,L}\big| \mathcal{G}_k  \big)-\mathbb{E}\big(\Psi_{f,L}\big| \mathcal{G}_{k-1}  \big)  \bigg)^2&
\leq 8\int_0^1\bigg[\mathbb{E} \bigg(\omega^2_{n_k}\big(\big\langle \delta_{n_k},  f'(H^\omega_L)\delta_{n_k}\big\rangle\big)^2\big|_{(\omega_{n_k}\to u\omega_{n_k})} \bigg)\bigg]du
\end{align*}
Use of the above in (\ref{fun-mrt-dis}) will give
\begin{align}
\label{final-inq-imp}
\mathbb{E}\big(\big| X_{f,L} \big| ^2 \big)&\leq 8\frac{1}{|\Lambda_L|}\sum_{k=1}^{(2L+1)^d}\int_0^1\bigg[\mathbb{E} \bigg(\omega^2_{n_k}\big(\big\langle \delta_{n_k},  f'(H^\omega_L)\delta_{n_k}\big\rangle\big)^2\big|_{(\omega_{n_k}\to u\omega_{n_k})} \bigg)\bigg]du\nonumber\\
& =8 \frac{1}{|\Lambda_L|}\sum_{n\in\Lambda_L}\int_0^1\bigg[\mathbb{E} \bigg(\omega^2_{n}\big(\big\langle \delta_{n},  f'(H^\omega_L)\delta_{n}\big\rangle\big)^2\big|_{(\omega_{n}\to u\omega_{n})} \bigg)\bigg]du\nonumber\\
& \leq8\frac{1}{|\Lambda_L|} \sum_{n\in\Lambda_L}\int_0^1\bigg[\mathbb{E} \bigg(\omega^2_{n}\big\langle \delta_{n},  \big|f'(H^\omega_L)\big|^2\delta_{n}\big\rangle\big|_{(\omega_{n}\to u\omega_{n})} \bigg)\bigg]du\nonumber\\
&= 8\int \big| f'(x) \big|^2d\bar{\nu}_L(x).
\end{align}
In the third inequality above, we have used the fact that for any self-adjoint operator $A$ on a Hilbert space $\mathcal{H}$ it is always true that $\langle \psi,  A\psi\rangle^2\leq \| \psi \|^2 \langle \psi,  A^2\psi\rangle~\forall~\psi\in\mathcal{H}$. 
Now taking $\limsup$ (w.r.t $L$) on both sides of the above (\ref{final-inq-imp}) together with Lemma \ref{weak-con} (in appendix) for $p=1$ will give the result.
\end{proof}
\begin{rem}
\label{rs-fr-chg-int}
In the above, all the changes of order in the integration are valid because, for $f\in C^1_P(\mathbb{R})$, we have $|f'(x)|\leq P(x)$ and now using Cauchy$-$Schwarz inequality together with (\ref{mono}),  (\ref{mono-fn}) and the Hypothesis \ref{hp2} one can show that 
$$\int_0^1\bigg[\mathbb{E}\bigg(\bigg|\omega_n^p\big\langle \delta_n, f'(H^\omega_L)\delta_n\big\rangle\big|_{(\omega_n\to u\omega_n)}  \bigg|\bigg)  \bigg]du<\infty,~p\in\mathbb{N}\cup\{0\}~~\text{for}~~\Lambda_L\subseteq \mathbb{Z}^d.$$
\end{rem}
\noindent 
Now we want to understand the properties of the limiting variance $\sigma^2_f$ for the test function $f\in C^1_P(\mathbb{R})$, and we start with the case when $f$ is a polynomial. 
 If we choose a polynomial as the test function in the definition (\ref{sm-rv-diff}), the limit in (\ref{df-lm-vr}) will exist, and it can be derived from the result of \cite{GYWJ}.
\begin{prop}
\label{poly-vr}
Let $P(x)=\displaystyle\sum_{k=0}^N a_k x^k$ be real polynomial of degree $N\ge 1$.  Consider the random variable $X_{P,L}$ as in (\ref{sm-rv-diff}) and $\sigma^2_P$ is the limiting variance as in the Theorem \ref{main}.  Then $\sigma^2_P=\displaystyle\lim_{L\to\infty}\mathbb{E}\big(\big| X_{P,L} \big|^2 \big)$.
\end{prop}
\begin{proof}
Using the \cite[equation (4.1)]{GYWJ} we get
\begin{equation}
\label{pl-v-1}
\sigma^2_P=\lim_{L\to\infty}\frac{1}{|\Lambda_L|}\text{Var}\bigg(\sum_{k=0}^Na_kT_k \bigg).
\end{equation}
The random variable $T_k$ is expressed in \cite[Definition 3.1]{GYWJ}.  In the proof of the \cite[Proposition 3.2]{GYWJ} it is shown that $\text{Var}\bigg( T_k-Tr\big(H^\omega_L\big)^k\bigg)=o(L^d)$.  Since $P$ is a fixed polynomial, therefore it is also immediate that
\begin{equation}
\label{pl-v-2}
\text{Var}\bigg(\sum_{k=0}^Na_kT_k-Tr\big(P\big(H^\omega_L\big)\big)\bigg)=o(L^d).
\end{equation}
Using Minkowski inequality,  we write
\begin{align}
\label{pl-v-3}
\begin{split}
\bigg(\frac{1}{|\Lambda_L|}\text{Var}\bigg(Tr\big(P\big(H^\omega_L\big)\big)\bigg)\bigg)^{\frac{1}{2}}&\leq \bigg(\text{Var}\bigg(\frac{1}{|\Lambda_L|}\sum_{k=0}^Na_kT_k-Tr\big(P\big(H^\omega_L\big)\big)\bigg)  \bigg)^{\frac{1}{2}}\nonumber\\
&\qquad\qquad +\bigg( \frac{1}{|\Lambda_L|}\text{Var}\bigg(\sum_{k=0}^Na_kT_k \bigg) \bigg)^{\frac{1}{2}}\\
\text{and}~\bigg( \frac{1}{|\Lambda_L|}\text{Var}\bigg(\sum_{k=0}^Na_kT_k \bigg) \bigg)^{\frac{1}{2}}
&\leq \bigg(\text{Var}\bigg(\frac{1}{|\Lambda_L|}\sum_{k=0}^Na_kT_k-Tr\big(P\big(H^\omega_L\big)\big)\bigg)  \bigg)^{\frac{1}{2}}\nonumber\\
&\qquad\qquad +\bigg(\frac{1}{|\Lambda_L|}\text{Var}\bigg(Tr\big(P\big(H^\omega_L\big)\big)\bigg)\bigg)^{\frac{1}{2}}.
\end{split}
\end{align}
The above two inequalities will give
\begin{align*}
&\bigg| \bigg( \frac{1}{|\Lambda_L|}\text{Var}\bigg(\sum_{k=0}^Na_kT_k \bigg) \bigg)^{\frac{1}{2}}-\bigg(\frac{1}{|\Lambda_L|}\text{Var}\bigg(Tr\big(P\big(H^\omega_L\big)\big)\bigg)\bigg)^{\frac{1}{2}} \bigg|\nonumber\\
&\qquad\qquad\qquad\qquad\qquad\qquad\leq \bigg(\text{Var}\bigg(\frac{1}{|\Lambda_L|}\sum_{k=0}^Na_kT_k-Tr\big(P\big(H^\omega_L\big)\big)\bigg)  \bigg)^{\frac{1}{2}}
\end{align*}
So we get from the above and (\ref{pl-v-2}) that
\begin{equation}
\label{pl-v-5}
\lim_{L\to\infty}\bigg| \bigg( \frac{1}{|\Lambda_L|}\text{Var}\bigg(\sum_{k=0}^Na_kT_k \bigg) \bigg)^{\frac{1}{2}}-\bigg(\frac{1}{|\Lambda_L|}\text{Var}\bigg(Tr\big(P\big(H^\omega_L\big)\big)\bigg)\bigg)^{\frac{1}{2}} \bigg|=0.
\end{equation}
Now using above and the limit (\ref{pl-v-1}) together with the definition (\ref{sm-rv-diff}) we get
$$\sigma^2_P=\displaystyle\lim_{L\to\infty}\mathbb{E}\big(\big| X_{P,L} \big|^2 \big).$$ 
Hence, the Proposition.
\end{proof}
\noindent Let $f\in C_{P}^1(\mathbb{R})$, then $f'\in L^2(\bar{\nu})$ (see Lemma \ref{weak-con} for $p=1$) therefore using the Corollary \ref{pol-dens} (in Appendix) we have a sequence of polynomials $\{P_k\}_{k=1}^\infty$ such that 
\begin{equation}
\label{poly-apprx-spect}
\| f'-P_k \|_{L^2(\bar{\nu})} \to 0 ~~\text{as}~~k\to\infty.
\end{equation} 
Denote the polynomial $Q_k$ as the primitive of $P_k$, i.e. $Q'_k=P_k$.  Now define the random variable
\begin{align}
\label{prim-rv}
X_{Q_k,L}(\omega)&= \frac{1}{|\Lambda_L|^\frac{1}{2}}\sum_{n\in\Lambda_L} \bigg(\langle \delta_n, Q_k(H^\omega_L)\delta_n\rangle - 
\mathbb{E}\big(\langle \delta_n, Q_k(H^\omega_L)\delta_n\rangle  \big)\bigg).
\end{align}
Now it is clear from the definition of $X_{f,L}$ (see (\ref{sm-rv-diff}) ) is that
\begin{equation}
\label{dif-ran-der}
X_{f,L}-X_{Q_k,L}=X_{(f-Q_k), L}.
\end{equation}
Under the Hypothesis \ref{hp2}, the SSD has all the moments; therefore, if we use the Proposition \ref{poly-vr} for the polynomial $P=Q_k$ we get
\begin{align}
\label{var-prim}
\sigma^2_k:=\sigma^2_{Q_k}=\lim_{L\to\infty}\mathbb{E}\bigg(\big| X_{Q_k,L}\big| ^2 \bigg)<\infty.
\end{align}
Now we will use the proposition \ref{fin-diff} to find the limit of $\sigma^2_k$ as $k\to\infty.$
\begin{lem}
\label{var-con-prim}
Let $f\in C^1_P(\mathbb{R})$, also consider $\sigma^2_f$ and $\sigma^2_k$ as given in (\ref{df-lm-vr}) and (\ref{var-prim}) respectively, then under the Hypothesis \ref{hp2} we have
\begin{equation}
\label{prim-var-con}
\lim_{k\to\infty}\sigma^2_k=\sigma^2_f.
\end{equation}
\end{lem}
\begin{proof}
Using Minkowski inequality and (\ref{dif-ran-der}) we write
\begin{align*}
\bigg(\mathbb{E}\bigg(\big| X_{f,L}\big| ^2 \bigg)\bigg)^{\frac{1}{2}}&\leq \bigg(\mathbb{E}\bigg(\big| X_{f,L}-X_{Q_k,L}\big| ^2 \bigg)\bigg)^{\frac{1}{2}}
+\bigg(\mathbb{E}\bigg(\big| X_{Q_k,L}\big| ^2 \bigg)\bigg)^{\frac{1}{2}}\\
  &=\bigg(\mathbb{E}\bigg(\big| X_{(f-Q_k),L}\big| ^2 \bigg)\bigg)^{\frac{1}{2}}+\bigg(\mathbb{E}\bigg(\big| X_{Q_k,L}\big| ^2 \bigg)\bigg)^{\frac{1}{2}}\\
  &\leq\sqrt{8} \bigg(\int\big| f'-P_k \big|^2d\bar{\nu}_L\bigg)^{\frac{1}{2}}+\bigg(\mathbb{E}\bigg(\big| X_{Q_k,L}\big| ^2 \bigg)\bigg)^{\frac{1}{2}},~~Q'_k=P_k.
\end{align*}
In the last line above, we have used the inequality (\ref{final-inq-imp}).  Now taking $\limsup$ (w.r.t $L$) of both side of the above and using Lemma \ref{weak-con} for $p=1$ we get
\begin{equation}
\label{one-inq}
\sigma_f\leq \sqrt{8} \| f'-P_k \|_{L^2(\bar{\nu})}+\sigma_k.
\end{equation}
Similarly, as above, the use of Minkowski inequality with the random variable $X_{Q_k,L}=\big(X_{Q_k,L}-X_{f_k,L}  \big)+X_{f,L}$ will give 
\begin{equation}
\label{two-inq}
\sigma_k\leq  \sqrt{8} \| f'-P_k \|_{L^2(\bar{\nu})}+\sigma_f.
\end{equation}
Two inequalities (\ref{one-inq}) and (\ref{two-inq}) will ensure $|\sigma_k-\sigma_f|\leq \sqrt{8} \| f'-P_k \|_{L^2(\bar{\nu})}$ and now the lemma will follow from the fact (\ref{poly-apprx-spect}).
\end{proof}
\noindent Now, we want to deduce a lower bound of $\sigma^2_P$, which will involve the expression of $P$ and to do that, we will use the properties of martingale.
Before that we define the families of $\sigma$-algebra $\big\{\mathcal{F}^m_k\}_{k\in\mathbb{Z}},~m=1,2,\cdots, d$ as 
\begin{align}
\begin{split}
\label{sig-al}
\mathcal{F}^m_k&=\sigma\bigg(\omega_n:n\in A^m_k\bigg),~A^m_k\subset \mathbb{Z}^d \\
 A^m_k&=\big\{(n_1,n_2,\cdots,n_d): n_m\leq k~ \text{and}~n_i\in\mathbb{Z}~\text{for}~i\neq m\big\}
 \end{split}
\end{align}
Here $\sigma\big(\omega_n : n\in \Lambda  \big),~\Lambda\subset\mathbb{Z}^d$ denote the $\sigma$-algebra generated by the collection of i.i.d real random variables $\big\{\omega_n : n\in \Lambda \big\}$. Now it is immediate from (\ref{sig-al}) that $A^m_k\subset A^m_{k+1}$ and $\mathcal{F}^m_k\subset \mathcal{F}^m_{k+1}~\forall~k\in\mathbb{Z}$ and $1\leq m \leq d$. 
 Also from the definition (\ref{restric}) of $H^\omega_L$ it is clear that the trace $\Psi_L(\omega):=Tr\big( P(H^\omega_L) \big)$ \big(as a random variable\big) 
 is measurable w.r.t the $\sigma$-algebra $\sigma\big(\omega_{(n_1,n_2,\cdots,  n_d)}: |n_i|\leq L~\forall~i\big)$.  Actually $\Psi_L(\omega)$ as a function depends only on the variables $\big\{\omega_{(n_1,n_2,\cdots,  n_d)}: |n_i|\leq L~\forall~i\big)  \big\}$.
Now,  we define the conditional expectation as
 \begin{equation}
 \label{cn-ex-1}
 \Psi_{L,1}(\omega):=\mathbb{E}\big(\Psi_L(\omega)\big| \mathcal{F}^1_{1}\big) -\mathbb{E}\big( \Psi_L(\omega)\big| \mathcal{F}^1_{0}\big).
 \end{equation}
For $2\leq \ell\leq d$ the random variable (conditional expectation) $\Psi_{L,1,2,\cdots, \ell}(\omega)$ is defined through the recursive relation
\begin{align}
\label{recur-cond}
\Psi_{L,1, 2,\cdots\ell}(\omega):=\mathbb{E}\big(\Psi_{L,1,2,\cdots, (\ell-1)}(\omega)\big| \mathcal{F}^\ell_{1}\big) -
\mathbb{E}\big( \Psi_{L,1,2, \cdots,(\ell-1)}(\omega)\big| \mathcal{F}^\ell_{0}\big).
\end{align}
We denote $\displaystyle \prod_{i=1}^\ell\mathcal{F}^{m_i}_{k_i}, (1\leq \ell\leq d)$ to be the $\sigma$-algebra generated by the collection of i.i.d random variables $\big\{\omega_n:n\in \displaystyle \cap _{i=1}^\ell A^{m_i}_{k_i} \big\}$, see (\ref{sig-al}) for the definition of the set $A^{m_i}_{k_i}\subset \mathbb{Z}^d$,  here $1\leq m_i\leq d$ and $k_i\in\mathbb{Z}$.
\begin{lem}
\label{st-pos-var}
Let $P$ be a real polynomial of degree $p\ge 1$ and consider $H^\omega_L$ and $\Psi_{L,1,2,\cdots,d}(\omega)$ as in (\ref{restric}) and (\ref{recur-cond}), then under the Hypothesis \ref{hp} we have 
\begin{equation}
\label{fnl-vr}
\sigma^2_P\ge\limsup_{L\to\infty}\mathbb{E}\big( \Psi^2_{L,1,2,\cdots,d}(\omega) \big)
\end{equation}
\end{lem}
\begin{proof}
Lets denote $\Psi_L(\omega)=Tr\big( P(H^\omega_L) \big)$ then the Proposition \ref{poly-vr} will give 
\begin{equation}
\label{us-fl}
\sigma^2_P=\lim_{L\to\infty}\frac{1}{|\Lambda_L|}\text{Var}\bigg(Tr\big(P\big(H^\omega_L\big) \big) \bigg).
\end{equation}
Since $\Psi_L(\omega)$ is depends only on $\{\omega_{(n_1,n_2,\cdots,n_d)}:|n_i|\leq L\}$ therefore using (\ref{sig-al}) we write
\begin{equation}
\label{mrt}
\mathbb{E}\big(\Psi_L  \big)=\mathbb{E}\big(\Psi_L\big| \mathcal{F}^1_{-(L+1)}  \big)~~\text{and}~~\Psi_L (\omega)=\mathbb{E}\big(\Psi_L\big| \mathcal{F}^1_{L}  \big).
\end{equation}
It can also be easily verified that the sequence of the conditional expectations $\big\{ \mathbb{E}\big( \Psi_L\big| \mathcal{F}^1_{k}\big) \big\}_{k=-L}^L$ 
is a martingale (Doob) w.r.t the filtration $\big\{\mathcal{F}^1_k\big\}_{k\in\mathbb{Z}}$.  Now we write the difference between $\Psi_L(\omega)$ and its expectation
$ \mathbb{E}\big(\Psi_L  \big)$ as a sum of the martingale differences
\begin{equation}
\label{mrt-dif}
\Psi_L(\omega)-\mathbb{E}\big(\Psi_L  \big)=\sum_{k=-L}^L\bigg(\mathbb{E}\big( \Psi_L(\omega)\big| \mathcal{F}^1_{k}\big) -\mathbb{E}\big( \Psi_L(\omega)\big| \mathcal{F}^1_{k-1}\big)\bigg).
\end{equation}
Using the formula (\ref{tr-fr}) (in appendix), the derivative of the trace, and the chain rule (of differentiation), we can write each martingale difference inside the above sum as
\begin{align}
\label{dis-same}
&\mathbb{E}\big( \Psi_L(\omega)\big| \mathcal{F}^1_{k}\big) -\mathbb{E}\big( \Psi_L(\omega)\big| \mathcal{F}^1_{(k-1)}\big),
~~~\text{here}~~\omega=(\omega_n)_{n\in\mathbb{Z}^d}\nonumber\\
&\qquad=\mathbb{E}\big( \Psi_L(\omega)\big| \mathcal{F}^1_{k}\big) -
\mathbb{E}\big( \Psi_L\big(\omega: (\omega_j=0)_{j\in A^1_k\setminus A^1_{k-1}}\big)\big| \mathcal{F}^1_{k}\big)\nonumber\\
&\qquad \qquad-\bigg(  \mathbb{E}\big( \Psi_L(\omega)\big| \mathcal{F}^1_{k-1}\big) -
\mathbb{E}\big( \Psi_L\big(\omega: (\omega_j=0)_{j\in A^1_k\setminus A^1_{k-1}}\big)\big| \mathcal{F}^1_{k}\big)\bigg)\nonumber\\
&\qquad=\mathbb{E}\big( \Psi_L(\omega)\big| \mathcal{F}^1_{k}\big) -
\mathbb{E}\big( \Psi_L\big(\omega: (\omega_j=0)_{j\in A^1_k\setminus A^1_{k-1}}\big)\big| \mathcal{F}^1_{k}\big)\nonumber\\
&\qquad\qquad -\bigg(\mathbb{E}\big( \Psi_L(\omega)\big| \mathcal{F}^1_{k-1}\big) -
\mathbb{E}\big( \Psi_L\big(\omega: (\omega_j=0)_{j\in A^1_k\setminus A^1_{k-1}}\big)\big| \mathcal{F}^1_{k-1}\big)\bigg)\nonumber\\
&\qquad=\mathbb{E}\bigg(\int_0^1\frac{d}{du}\big( \Psi_L\big(\omega: (\omega_j\to u\omega_j)_{j\in A^1_k\setminus A^1_{k-1}}\big)\big)du\bigg| \mathcal{F}^1_{k}\bigg)\nonumber\\
&\qquad \qquad-\mathbb{E}\bigg(\int_0^1\frac{d}{du}\big( \Psi_L\big(\omega: (\omega_j\to u\omega_j)_{j\in A^1_k\setminus A^1_{k-1}}\big)\big)du\bigg| \mathcal{F}^1_{k-1}\bigg)\nonumber\\
&\qquad=\sum_{n\in A^1_k\setminus A^1_{k-1}} \mathbb{E}\bigg( \int_0^1\omega_n\langle \delta_n,  P'(H^\omega_L)\delta_n\rangle\big|_{(\omega_j\to u\omega_j)_{j\in A^1_k\setminus A^1_{k-1}}} du \bigg| \mathcal{F}^1_{k} \bigg)\nonumber\\
&\qquad\qquad-
\sum_{n\in A^1_k\setminus A^1_{k-1}} \mathbb{E}\bigg( \int_0^1\omega_n\langle \delta_n,  P'(H^\omega_L)\delta_n\rangle\big|_{(\omega_j\to u\omega_j)_{j\in A^1_k\setminus A^1_{k-1}}}du \bigg| \mathcal{F}^1_{k-1} \bigg)\nonumber\\
&\qquad =\sum_{n\in A^1_k\setminus A^1_{k-1}}\int_0^1\bigg[ \mathbb{E}\bigg(\omega_n\langle \delta_n,  P'(H^\omega_L)\delta_n\rangle\big|_{(\omega_j\to u\omega_j)_{j\in A^1_k\setminus A^1_{k-1}}} \bigg| \mathcal{F}^1_{k} \bigg)\nonumber\\
&\qquad\qquad -\int_{_{\mathbb{R}^{\big|A^1_k\setminus A^1_{k-1}\big|}}} \bigg(\mathbb{E}\bigg( \omega_n\langle \delta_n,  P'(H^\omega_L)\delta_n\rangle\big|_{(\omega_j\to u\omega_j)_{j\in A^1_k\setminus A^1_{k-1}}} \bigg| \mathcal{F}^1_{k} \bigg)\bigg)\nonumber\\
&\qquad\qquad\qquad\qquad\qquad\qquad \qquad \times \prod_{j\in A^1_k\setminus A^1_{k-1}}d\mu(\omega_j)\bigg]du.
\end{align}
Since $P'(H^\omega_L)$ is function of $(\omega_n)_{n\in\Lambda_L}$ only so the above sum is actually over all
$n\in \Lambda_L\cap\big(A^1_k\setminus A^1_{k-1}  \big)=\big\{(k, n_2,n_3,\cdots,n_d): |n_i|\leq L,~i=2,3,\cdots, d, \big\}$, $|k|\in\Lambda_L$ is fixed,  see the definition (\ref{sig-al}) and (\ref{box}).  Also, the changes in the order of integrations above are valid because of Remark \ref{reas-ch-int}, below.\\
Now $P'$ is a polynomial of degree $p-1$ and $\{\omega_n\}_{n\in\mathbb{Z}^d}$ are i.i.d random variables, therefore using the definition (\ref{restric}) and the above (\ref{dis-same}) we get that the collection of conditional expectations $\bigg\{\mathbb{E}\big( \Psi_L(\omega)\big| \mathcal{F}^1_{k}\big) -\mathbb{E}\big( \Psi_L(\omega)\big| \mathcal{F}^1_{(k-1)}\big) \bigg\}_{k=-( L-p)}^{L-p}$ has same distribution (may not independent).  In particular for $-(L-p)\leq k\leq L-p$, we have
\begin{equation}
\label{lw-bdd}
\begin{split}
&\mathbb{E}\bigg( \mathbb{E}\big( \Psi_L(\omega)\big| \mathcal{F}^1_{k}\big) -\mathbb{E}\big( \Psi_L(\omega)\big| \mathcal{F}^1_{(k-1)}\big) \bigg)^2=\mathbb{E}\bigg(\Psi^2_{L,1}(\omega)  \bigg).\\
& \text{Here the random variable} ~\Psi_{L,1}(\omega):=\mathbb{E}\big(\Psi_L(\omega)\big| \mathcal{F}^1_{1}\big) -\mathbb{E}\big( \Psi_L(\omega)\big| \mathcal{F}^1_{0}\big).
\end{split}
\end{equation}
Since the covariance between any two distinct elements from a martingale (Doob) difference sequence is always zero,  see Proposition \ref{or-db-dif-mart}, in the appendix, therefore using the (\ref{mrt-dif}) and (\ref{lw-bdd}), we estimate a lower bound of the variance of the random variable $\Psi_L(\omega)$ as
\begin{align}
\label{lbd-1}
\text{Var}(\Psi_L)&=\mathbb{E}\bigg(  \Psi_L(\omega)-\mathbb{E}\big(\Psi_L  \big)\bigg)^2\nonumber\\
&=\sum_{k=-L}^L\mathbb{E}\bigg(\mathbb{E}\big( \Psi_L(\omega)\big| \mathcal{F}^1_{k}\big) -\mathbb{E}\big( \Psi_L(\omega)\big| \mathcal{F}^1_{k-1}\big)\bigg)^2\nonumber\\
&\ge \sum_{k=-(L-p)}^{L-p}\mathbb{E}\bigg(\mathbb{E}\big( \Psi_L(\omega)\big| \mathcal{F}^1_{k}\big) -\mathbb{E}\big( \Psi_L(\omega)\big| \mathcal{F}^1_{k-1}\big)\bigg)^2\nonumber\\
&=(2L-2p+1)~\mathbb{E}\bigg(\Psi^2_{L,1}(\omega)  \bigg).
\end{align}
Again we define the Doob martingale $\big\{\mathbb{E}\big( \Psi_{L,1}\big|\mathcal{F}^2_k \big) \big\}_{k\in\mathbb{Z}}$ w.r.t the filtration $\big\{\mathcal{F}^2_k  \big\}_{k\in\mathbb{Z}}$ as defined by (\ref{sig-al}) and it will also give $\mathbb{E}\big(\Psi_{L,1} \big)=0$ (Law of total expectation).  Now, the same (types) method which we have used to get (\ref{lbd-1}) will give the lower bound of the variance of $\Psi_{L,1}$ as
\begin{align}
\label{lbd-2}
\begin{split}
& \text{Var}\big(\Psi_{L,1}  \big)\ge (2L-2p+1)\mathbb{E}\bigg(\Psi^2_{L,1,2}(\omega)  \bigg).\\
& \text{Here} ~\Psi_{L,1, 2}(\omega):=\mathbb{E}\big(\Psi_{L,1}(\omega)\big| \mathcal{F}^2_{1}\big) -\mathbb{E}\big( \Psi_{L,1}(\omega)\big| \mathcal{F}^2_{0}\big).
\end{split}
\end{align}
So the lower bound of the variance in (\ref{lbd-1}) will transform into a new one as
\begin{align}
\label{lbdnw-2}
\text{Var}\big(\Psi_L  \big)\ge (2L-2p+1)^2~ \mathbb{E}\bigg(\Psi^2_{L,1,2}(\omega)  \bigg).
\end{align}
Using the (\ref{lw-bdd}), (\ref{lbd-2}) and the Proposition \ref{con-ex-id} (in the appendix), the random variable $\Psi_{L,1, 2}$ can also be written as a sum of conditional expectations of the random variable $\Psi_{L}$ as
\begin{align}
\label{suc-rv}
\Psi_{L,1, 2}(\omega)&=\mathbb{E}\big(\Psi_{L,1}(\omega)\big| \mathcal{F}^2_{1}\big) -\mathbb{E}\big( \Psi_{L,1}(\omega)\big| \mathcal{F}^2_{0}\big)\nonumber\\
&=\bigg(\mathbb{E}\big(\Psi_{L}(\omega)\big| \mathcal{F}^1_1\mathcal{F}^2_{1}\big)-\mathbb{E}\big(\Psi_{L}(\omega)\big| \mathcal{F}^1_1\mathcal{F}^2_{0}\big)\bigg)\nonumber\\
& \qquad \qquad -\bigg(\mathbb{E}\big(\Psi_{L}(\omega)\big| \mathcal{F}^1_0\mathcal{F}^2_{1}\big)-\mathbb{E}\big(\Psi_{L}(\omega)\big| \mathcal{F}^1_0\mathcal{F}^2_{0}\big)  \bigg).
\end{align}
In the above we denote $\displaystyle \prod_{i=1}^\ell\mathcal{F}^{m_i}_{k_i}, (1\leq \ell\leq d)$ to be the $\sigma$-algebra generated by the collection of i.i.d random variables $\big\{\omega_n:n\in \displaystyle \cap _{i=1}^\ell A^{m_i}_{k_i} \big\}$,  see the discussion below
the (\ref{recur-cond}).\\
Now we will repeat all the steps (\ref{lbd-2})-(\ref{suc-rv}) for another $d-2$ times to get
\begin{align}
\label{lw-bdd-ls}
\text{Var}\big(\Psi_L  \big)\ge (2L-2p+1)^d~ \mathbb{E}\bigg(\Psi^2_{L,1,2,\cdots, d}(\omega)  \bigg).
\end{align}
The random variable $\Psi_{L,1,2,\cdots, d}(\omega)$ is defined through the recursive relation (\ref{recur-cond}),  as we did in (\ref{lw-bdd}) and (\ref{lbd-2}) for $\Psi_{L,1}(\omega)$ and $\Psi_{L,1, 2}(\omega)$, respectively
As $\big| \Lambda_L\big|=(2L+1)^d$ therefore using (\ref{us-fl}) in the (\ref{lw-bdd-ls}) we get the lower bound of the limiting variance $\sigma^2_P$  as
\begin{align}
\sigma^2_P=\lim_{L\to\infty}\frac{1}{|\Lambda_L|}\text{Var}\big(\Psi_L  \big)\ge\limsup_{L\to\infty}\mathbb{E}\big( \Psi^2_{L,1,2,\cdots,d}(\omega) \big).
\end{align}
 Hence, the Proposition.
\end{proof}
\begin{rem}
\label{reas-ch-int}
In the above, all changes of order in the integration are valid because, for any $B\subset\Lambda_L$ using (\ref{mono}) and the Hypothesis \ref{hp}, it immediately follows that 
$$\int_0^1\bigg[\mathbb{E}\bigg(\bigg|\omega_n\big\langle \delta_n,  P'(H^\omega_L)\delta_n\big\rangle \big|_{(\omega_j\to u\omega_j)_{j\in B}}  \bigg|  \bigg)   \bigg]du<\infty.$$
\end{rem}
\begin{rem}
As it is done in (\ref{suc-rv}), the random variable $\Psi_{L,1,2,\cdots,d}(\omega)$ can be written as the sum of conditional expectations of $\Psi_L(\omega)$.  So we write
\begin{align}
\label{dth-step}
\Psi_{L,1,2,\cdots,d}(\omega)&=\bigg(\mathbb{E}\big(\Psi_{L}(\omega)\big| \mathcal{F}^1_1\mathcal{F}^2_{1}\cdots
\mathcal{F}^{d-1}_1\mathcal{F}^d_1\big)-
\mathbb{E}\big(\Psi_{L}(\omega)\big| \mathcal{F}^1_1\mathcal{F}^2_{1}\cdots
\mathcal{F}^{d-1}_1\mathcal{F}^d_0\big)\bigg)\nonumber\\
&\qquad \qquad +\Phi_{L,d}(\omega).
\end{align}
In the above the random variable $\Phi_{L,d}(\omega)$ denote the sum of $(2d-2)$ many conditional expectations of the form
$\mathbb{E}\bigg(\Psi_{L}(\omega)\big| \displaystyle\prod_{i=1}^d\mathcal{F}^i_{k_i}\bigg)$ or $\mathbb{E}\bigg(-\Psi_{L}(\omega)\big| \displaystyle\prod_{i=1}^d\mathcal{F}^i_{k_i}\bigg)$ where $k_i\in \{0,1\}$ and $\big(k_1,k_2,\cdots,k_{d-1},k_d  \big)\neq (1,1,\cdots,1,1)$.  Therefore its clear from the definition (\ref{sig-al}) of the $\sigma$-algebra $\mathcal{F}^m_k$ that the two random variables $\Phi_{L,d}(\omega)$ and $\omega_{(1,1,\cdots,1,1)}$ are independent for all $L\ge 1$.
\end{rem}
\noindent 
For simplicity of the calculation, first, we will assume that the $\mu$ (SSD) is a compactly supported probability measure on $\mathbb{R}$.
Now, using the polynomial approximation of continuous function on a compact set (Weierstrass approximation theorem), we can show the limiting variance $\sigma^2_f$ is a positive (strictly) quantity whenever $f$ is a strictly monotone function on a deterministic open interval $I$ where $\sigma(H^\omega)\subseteq I$,  the spectrum $\sigma(H^\omega)$ is deterministic compact set a.e $\omega$.  
\begin{lem}
\label{pos-vr-mn-fn}
Assume the single site distribution (SSD) $\mu$ is a compactly supported probability measure on $\mathbb{R}$.  Let $f\in C_P^1(\mathbb{R})$ and it is strictly monotone function on an open interval $I$ (deterministic) where $\sigma(H^\omega)\subseteq I~a.e~\omega$ then the limiting variance $\sigma^2_f$ \big(as in (\ref{df-lm-vr})\big) is strictly positive.
\end{lem}
\begin{proof}
Since the probability measure (SSD) $\mu$ is compactly supported on $\mathbb{R}$ then there always exist a non-random closed bounded interval $J\subset I$ such that the spectrum $\sigma\big(H^\omega|_{(\omega_n\to u\omega_n)}\big)\subseteq J\subset I~a.e~\omega$ for all $0\leq u \leq 1$ and $n\in\mathbb{Z}^d$. Therefore the support of the  finite measure $\bar{\nu}$ \big(as in (\ref{full-ldosm})\big) is contained in $J$ i.e  $supp(\bar{\nu})\subseteq J$.  Let $\{P_k\}_k$ is a sequence of polynomials converges uniformly to $f'$ on the closed bounded interval $J$, i.e
\begin{equation}
\label{unif-app}
\big\|P_k-f' \big\|_{\infty}\to 0~~\text{also we have}~~\big\| P_k-f' \big\|_{L^2(\bar{\nu})}\to 0 ~\text{as}~k\to\infty.
\end{equation}
Lets denote $Q_k'=P_k$ then using the same notations as in Lemma \ref{var-con-prim} we get
\begin{equation}
\label{pl-ap-vr}
\lim_{k\to\infty} \sigma^2_k=\sigma^2_f,~~\text{here}~\sigma^2_k:=\sigma^2_{Q_k}=\lim_{L\to\infty}\mathbb{E}\bigg(\big| X_{L,Q_k} \big| ^2 \bigg).
\end{equation}
Now using the above and (\ref{fnl-vr}) for $P=Q_k$ we write
\begin{equation}
\label{vr-f-lw}
\sigma^2_f=\lim_{k\to\infty}\sigma^2_{Q_k}\ge \limsup_{k\to\infty}\limsup_{L\to\infty}
\mathbb{E}\big(\Psi^2_{k,L,1,2,\cdots,d}(\omega)  \big).
\end{equation}
In the above $\Psi_{k,L,1,2,\cdots,d}(\omega)$ same as $\Psi_{L,1,2,\cdots,d}(\omega)$ is given by the (\ref{dth-step}) when $\Psi_L(\omega):=\Psi_{k,L}(\omega)=Tr\big(Q_k(H^\omega_L)  \big)$, i.e we replace the polynomial $P$ by $Q_k$.  We can take the degree of the polynomial $P_k =Q_k'$ to be less or equal $k$, for example, Bernstein polynomial (see \cite{BR}).  Now using the Proposition \ref{doule-seq} (in appendix) in (\ref{vr-f-lw}) we get
\begin{equation}
\label{sb-seq-lw-bd}
\sigma^2_f\ge \limsup_{k\to\infty}\mathbb{E}\big(\Psi^2_{k,L_k,1,2,\cdots,d}(\omega)  \big)~~\text{and}~ ~L_k>k^2.
\end{equation}
The expression of $\Psi_{L,1,2,\cdots,d}(\omega)$ is given in (\ref{dth-step}) and now replacing the polynomial $P$ by $Q_k$ and $L$ by $L_k$ we write
\begin{equation}
\label{ag-us-mrt}
\Psi_{k,L_k,1,2,\cdots,d}(\omega)=\mathbb{E}\big(\Psi_{k,L_k}(\omega)\big| \mathcal{F}^1_1\mathcal{F}^2_{1}\cdots
\mathcal{F}^{d-1}_1\mathcal{F}^d_1\big)+V_{k,L_k}(\omega).
\end{equation}
In the above the random variable $V_{k,L_k}$ is given by
\begin{align}
\label{v_k-def}
V_{k,L_k}(\omega)&=-
\mathbb{E}\big(\Psi_{k,L_k}(\omega)\big| \mathcal{F}^1_1\mathcal{F}^2_{1}\cdots
\mathcal{F}^{d-1}_1\mathcal{F}^d_0\big)
+\Phi_{k,L_k,d}(\omega).
\end{align}
It is clear from the definition of the $\sigma$-algebra $\displaystyle \prod_{i=1}^d \mathcal{F}^i_{k_i}$,  here $k_i\in\{0,1\}$ and also 
$\big(k_1,k_2,\cdots, k_d \big)\neq(1,1,\cdots,1)$ that the random variable $V_{k,L_k}(\omega)$ is independent of the random variable $\omega_{(1,1,\cdots,1)}$ for large $L_k$, for more details we refer to the discussion just below the equation (\ref{dth-step}).\\
The trace $\Psi_{k,L_k}(\omega)=Tr\big(Q_k(H^\omega_{L_k})  \big)$ depends only on the random variables $\big\{\omega_n: |n|\leq L_k  \big\}$ (see \ref{restric}) and it will imply the function $\Psi_{k,L_k}$ measurable w.r.t the $\sigma$-algebra generated by the random variables $\big\{\omega_n  \big\}_{n\in\Lambda_{L_k}}$. \\
Let $\big\{ m_j \big\}_{j=1}^{(2L_k+1)^d}$ with $m_j<m_{j+1}$ is an enumeration of the elements of the finite cube $\Lambda_{L_k}$. Now define a filtrations of $\sigma$-algebra $\big\{\mathcal{D}_j \big\}_{j=1}^{N_{L_k}}$ as
\begin{equation}
\label{agn-sig-algb}
\mathcal{D}_j =\sigma \big(\omega_m: 1\leq m \leq m_j\big)~\text{and also we have }~\mathcal{D}_j\subset \mathcal{D}_{j+1}.
\end{equation}
For large enough $L_k$ we always have $(1,1,\cdots,1)\in\Lambda_{L_k}$ therefore w.l.o.g we can assume $m_{1}$ is the vector $
(1,1,\cdots,1)$.  \\
We denote  $\mathcal{D}_0$ as the trivial $\sigma$-algebra, i.e., it consists of empty and total space.
Since $\mathbb{E}\bigg (\Psi_{k,L_k,1,2,\cdots,d}(\omega) \bigg)=0$, therefore using Proposition \ref{or-db-dif-mart} we write the variance of $\Psi_{k,L_k,1,2,\cdots,d}(\omega)$ as
\begin{align}
\label{wr-vr-ag}
\mathbb{E}\big(\Psi^2_{k,L_k,1,2,\cdots,d}(\omega) \big)&=
\sum_{j=1}^{(2L_k+1)^d}\mathbb{E}\bigg[\bigg(\mathbb{E}\big(\Psi_{k,L_k,1,2,\cdots,d}(\omega) \big| \mathcal{D}_j \big)\nonumber\\
&\qquad \qquad \qquad-\mathbb{E}\big(\Psi_{k,L_k,1,2,\cdots,d}(\omega) \big| \mathcal{D}_{j-1}\big)\bigg)^2\bigg]\nonumber\\
&\geq \mathbb{E}\bigg[\bigg(\mathbb{E}\big(\Psi_{k,L_k,1,2,\cdots,d}(\omega) \big| \mathcal{D}_{1} \big)\nonumber\\
&\qquad \qquad \qquad-\mathbb{E}\big(\Psi_{k,L_k,1,2,\cdots,d}(\omega) \big| \mathcal{D}_{0}\big)\bigg)^2\bigg]\nonumber\\
&\geq \mathbb{E}\bigg[\bigg(\mathbb{E}\big(\Psi_{k,L_k,1,2,\cdots,d}(\omega) \big| \mathcal{D}_{1} \big)\bigg)^2\bigg].
\end{align}
In the last line of above, we have used the fact that for a trivial $\sigma$-algebra $\mathcal{D}_0$, we always have $\mathbb{E}\big(\Psi_{k,L_k,1,2,\cdots,d}(\omega) \big| \mathcal{D}_{0}\big)=
\mathbb{E}\big(\Psi_{k,L_k,1,2,\cdots,d}(\omega) \big)=0$.  
Again, using this very same fact and the expression (\ref{ag-us-mrt}), we also have
\begin{align}
\label{ag-dr-frml}
&\mathbb{E}\big(\Psi_{k,L_k,1,2,\cdots,d}(\omega) \big| \mathcal{D}_{1} \big)\nonumber\\
&\qquad \qquad  =\mathbb{E}\big(\Psi_{k,L_k,1,2,\cdots,d}(\omega) \big| \mathcal{D}_{1} \big)-\mathbb{E}\big(\Psi_{k,L_k,1,2,\cdots,d}(\omega) \big| \mathcal{D}_{0} \big)\nonumber\\
&\qquad \qquad =\mathbb{E}\bigg[\bigg(\mathbb{E}\big(\Psi_{k,L_k}(\omega)\big| \mathcal{F}^1_1\mathcal{F}^2_{1}\cdots
\mathcal{F}^{d-1}_1\mathcal{F}^d_1\big)+V_{k,L_k}(\omega)\bigg)\bigg| \mathcal{D}_1 \bigg]\nonumber\\
&\qquad \qquad \qquad \qquad -\mathbb{E}\bigg[\bigg(\mathbb{E}\big(\Psi_{k,L_k}(\omega)\big| \mathcal{F}^1_1\mathcal{F}^2_{1}\cdots
\mathcal{F}^{d-1}_1\mathcal{F}^d_1\big)+V_{k,L_k}(\omega)\bigg)\bigg| \mathcal{D}_0 \bigg]\nonumber\\
&\qquad \qquad=\mathbb{E}\bigg[\bigg(\mathbb{E}\big(\Psi_{k,L_k}(\omega)\big| \mathcal{F}^1_1\mathcal{F}^2_{1}\cdots
\mathcal{F}^{d-1}_1\mathcal{F}^d_1\big)\bigg)\bigg| \mathcal{D}_1 \bigg]\nonumber\\
&\qquad\qquad \qquad \qquad-\mathbb{E}\bigg[\bigg(\mathbb{E}\big(\Psi_{k,L_k}(\omega)\big| \mathcal{F}^1_1\mathcal{F}^2_{1}\cdots
\mathcal{F}^{d-1}_1\mathcal{F}^d_1\big)\bigg)\bigg| \mathcal{D}_0 \bigg]\nonumber\\
&\qquad \qquad =\mathbb{E}\big( \Psi_{k,L_k}(\omega)\big|\mathcal{D}_1 \big)-
\mathbb{E}\big( \Psi_{k,L_k}(\omega)\big|\mathcal{D}_0 \big)\nonumber\\
&\qquad \qquad =\mathbb{E}\big( \Psi_{k,L_k}(\omega)\big|\mathcal{D}_1 \big)-\mathbb{E}\big( \Psi_{k,L_k}(\omega)\big|_{\omega_{m_1}=0}\big|\mathcal{D}_1 \big)\nonumber\\
&\qquad\qquad \qquad \qquad-\bigg(\mathbb{E}\big( \Psi_{k,L_k}(\omega)\big|\mathcal{D}_0 \big)
-\mathbb{E}\big( \Psi_{k,L_k}(\omega)\big|_{\omega_{m_1}=0}\big|\mathcal{D}_1 \big)\bigg)
\nonumber\\
&\qquad \qquad =\mathbb{E}\big( \Psi_{k,L_k}(\omega)\big|\mathcal{D}_1 \big)-\mathbb{E}\big( \Psi_{k,L_k}(\omega)\big|_{\omega_{m_1}=0}\big|\mathcal{D}_1 \big)\nonumber\\
&\qquad\qquad \qquad \qquad-\bigg(\mathbb{E}\big( \Psi_{k,L_k}(\omega)\big|\mathcal{D}_0 \big)
-\mathbb{E}\big( \Psi_{k,L_k}(\omega)\big|_{\omega_{m_1}=0}\big|\mathcal{D}_0 \big)\bigg).
\end{align}
In the third equality above, we have used the fact that $V_{k,L_k}(\omega)$ is independent of $\mathcal{D}_1$,  $\sigma$-algebra generated by the single random variable, namely $\omega_{m_1}=\omega_{(1,1,\cdots,1)}$ and in the fourth equality we made use of the inclusion (of $\sigma$-algebra) $\mathcal{D}_0\subset \mathcal{D}_1\subset \mathcal{F}^1_1\mathcal{F}^2_{1}\cdots \mathcal{F}^{d-1}_1\mathcal{F}^d_1 $.\\
Since $\Psi_{k,L_k}(\omega)=Tr\big(Q_k\big( H^\omega_{L_k}\big)  \big)  \big),~Q_k'=P_k$, therefore using the derivative of the trace as in Lemma \ref{der-tr} (appendix) we write
\begin{align}
\label{sing-pertr-us}
&\Psi_{k,L_k}(\omega)-\Psi_{k,L_k}(\omega)\big|_{\omega_{m_1}=0}\nonumber\\
&\qquad \qquad =\int_0^1\frac{d}{du}\bigg(\Psi_{k,L_k}(\omega)\big|_{(\omega_{m_1}\to u\omega_{m_1})}\bigg)du\nonumber\\
&\qquad \qquad =\int_0^1\bigg(\omega_{m_1}\big\langle \delta_{m_1}, Q'_k\big( H^\omega_{L_k} \big)\delta_{m_1}  \big\rangle\big|_{(\omega_{m_1}\to u\omega_{m_1})}\bigg) du\nonumber\\
&\qquad \qquad =\int_0^1\bigg(\omega_{m_1}\big\langle \delta_{m_1}, Q'_k\big( H^\omega \big)\delta_{m_1}  \big\rangle\big|_{(\omega_{m_1}\to u\omega_{m_1})}\bigg) du\nonumber\\
&\qquad \qquad =\int_0^1\bigg(\omega_{m_1}\big\langle \delta_{m_1}, P_k\big( H^\omega \big)\delta_{m_1}  \big\rangle\big|_{(\omega_{m_1}\to u\omega_{m_1})}\bigg) du.
\end{align}
In the third line of the above, we used the fact that $Q'_k\big( H^\omega_{L_k})\delta_{m_1}=Q'_k\big( H^\omega \big)\delta_{m_1}$ for
$m_1\in\Lambda^{int}_{L_k,k},~m_1=(1,1,\cdots,1)$, here the degree of the polynomial $Q'_k$ is less or equal $k$, and we also have $L_k>k^2$.\\
Now using the above (\ref{sing-pertr-us}) in (\ref{ag-dr-frml}) we get
\begin{align}
\label{der-mrt-estm}
&\mathbb{E}\big(\Psi_{k,L_k,1,2,\cdots,d}(\omega) \big| \mathcal{D}_{1} \big)\nonumber\\
&\qquad \qquad = \mathbb{E}\bigg(\int_0^1\bigg(\omega_{m_1}\big\langle \delta_{m_1}, P_k\big( H^\omega \big)\delta_{m_1}  \big\rangle\big|_{(\omega_{m_1}\to u\omega_{m_1})}\bigg)du\bigg|\mathcal{D}_1\bigg)\nonumber\\ 
& \qquad \qquad \qquad -\mathbb{E}\bigg(\int_0^1\bigg(\omega_{m_1}\big\langle \delta_{m_1}, P_k\big( H^\omega \big)\delta_{m_1}  \big\rangle\big|_{(\omega_{m_1}\to u\omega_{m_1})}\bigg)du\bigg|\mathcal{D}_0\bigg)\nonumber\\
&\qquad \qquad = \mathbb{E}\bigg(\int_0^1\bigg(\omega_{m_1}\big\langle \delta_{m_1}, P_k\big( H^\omega \big)\delta_{m_1}  \big\rangle\big|_{(\omega_{m_1}\to u\omega_{m_1})}\bigg)du\bigg|\omega_{m_1}\bigg)\nonumber\\ 
& \qquad \qquad \qquad -\mathbb{E}\bigg(\int_0^1\bigg(\omega_{m_1}\big\langle \delta_{m_1}, P_k\big( H^\omega \big)\delta_{m_1}  \big\rangle\big|_{(\omega_{m_1}\to u\omega_{m_1})}\bigg)du\bigg).
\end{align}
In the last line of the above, we used the fact that $\mathcal{D}_1=\sigma(\omega_{m_1})$, $\sigma$-algebra generated by the random variable $\omega_{m_1}$  and $\mathcal{D}_0$ is the trivial $\sigma$-algebra consisting of full and empty space.\\
Since $\mu$ (SSD) is compactly supported so
$\{\omega_n\}_{n\in\mathbb{Z}^d}$ are i.i.d  bounded random variables. It has already been discussed at the beginning of the proof that for $0\leq u\leq1$ we always have $\sigma\bigg(H^\omega\big|_{(\omega_{m_1}\to u\omega_{m_1})} \bigg)\subseteq J$ a.e $\omega$, where $J$ is a closed and bounded interval (deterministic).  Now the uniform convergence of $\{P_k\}_k$ (polynomials) to $f'$ on the compact set 
$J$ together with (\ref{der-mrt-estm}) will give the existence of the limit
\begin{align}
\label{lim-exist}
&\lim_{k\to\infty}\mathbb{E}\big(\Psi_{k,L_k,1,2,\cdots,d}(\omega) \big| \mathcal{D}_{1} \big)\nonumber\\
&\qquad \qquad = \mathbb{E}\bigg(\int_0^1\bigg(\omega_{m_1}\big\langle \delta_{m_1}, f'\big( H^\omega \big)\delta_{m_1}  \big\rangle\big|_{(\omega_{m_1}\to u\omega_{m_1})}\bigg)du\bigg|\omega_{m_1}\bigg)\nonumber\\ 
& \qquad \qquad \qquad -\mathbb{E}\bigg(\int_0^1\bigg(\omega_{m_1}\big\langle \delta_{m_1}, f'\big( H^\omega \big)\delta_{m_1}  \big\rangle\big|_{(\omega_{m_1}\to u\omega_{m_1})}\bigg)du\bigg)\nonumber\\
&\qquad \qquad = \omega_{m_1}\int_0^1\mathbb{E}\bigg( \big\langle \delta_{m_1}, f'\big( H^\omega \big)\delta_{m_1}  \big\rangle\big|_{(\omega_{m_1}\to u\omega_{m_1})}\bigg|\omega_{m_1}\bigg)du\nonumber\\
&\qquad \qquad \qquad-\int_0^1\mathbb{E}\bigg(\omega_{m_1} \big\langle \delta_{m_1}, f'\big( H^\omega \big)\delta_{m_1}  \big\rangle\big|_{(\omega_{m_1}\to u\omega_{m_1})}\bigg)du.
\end{align}
Since $\{\omega_n\}_{n\in\mathbb{Z}^d}$ are bounded random variables and also $f'$ is continuous on the compact interval
$J$,  so in the last line of the above, we have used Fubini's theorem to change the order of the integration together with the property of conditional expectation, namely $\mathbb{E}\big(XY\big|X \big)=X \mathbb{E}\big(Y\big|X  \big)$.\\
Now, using the spectral measure of the self-adjoint operator 
$H^\omega|_{(\omega_{m_1}\to u \omega_{m_1})}$ at the vector $\delta_{m_1}$ we write the r.h.s of (\ref{lim-exist}) as
\begin{align}
\label{lim-to-non-zero}
&\omega_{m_1}\int_0^1\mathbb{E}\bigg( \big\langle \delta_{m_1}, f'\big( H^\omega \big)\delta_{m_1}  \big\rangle\big|_{(\omega_{m_1}\to u \omega_{m_1})}\bigg|\omega_{m_1}\bigg)du\nonumber \\
&\qquad \qquad \qquad \qquad \qquad-\int_0^1\mathbb{E}\bigg(\omega_{m_1} \big\langle \delta_{m_1}, f'\big( H^\omega \big)\delta_{m_1}  \big\rangle\big|_{(\omega_{m_1}\to u\omega_{m_1})}\bigg)du\nonumber\\
 & \qquad \qquad \qquad   =\omega_{m_1}\int f'(x)d\mu_{\omega_{m_1}}^\perp(x)
 -\mathbb{E}\bigg( \omega_{m_1}\int f'(x)d\mu_{\omega_{m_1}}^\perp(x) \bigg).
\end{align}
In the above the probability measure $\mu_{\omega_{m_1}}^\perp(\cdot)$ is given by
\begin{equation}
\label{perp-meas}
\mu_{\omega_{m_1}}^\perp(\cdot):=\int_0^1\mathbb{E}\bigg( \big\langle \delta_{m_1},  E_{H^\omega}(\cdot)\delta_{m_1}  \big\rangle\big|_{(\omega_{m_1}\to u \omega_{m_1})}\bigg|\omega_{m_1}\bigg)du.
\end{equation}
Here $\big\langle \delta_{m_1}, E_{ H^\omega}(\cdot)\delta_{m_1}  \big\rangle\big|_{(\omega_{m_1}\to u\omega_{m_1})}$ denote 
the spectral measure of the operator $H^\omega\big|_{(\omega_{\omega_{m_1}\to u \omega_{m_1} })}$ at the vector $\delta_{m_1}$.
Also it will follow from the result  \cite{ds} that the probability measure $\mu_{\omega_{m_1}}^\perp(\cdot)$ is non-degenerate $a.e~\omega_{m_1}$.\\
W.l.o.g let assume $f\in C^1_P(\mathbb{R})$ is strictly monotone increasing function on the open interval $I$, where $\sigma\big(H^\omega\big|_{(\omega_{m_1}\to u\omega_{m_1})}\big)\subseteq J\subset I$ for $0\leq u\leq 1$,  so $f'(x)>0$ on $J$,  where $J$ is a compact interval (deterministic), see the discussion at the beginning of the proof for the description of $J$.  
Therefore we have 
\begin{equation}
\label{innr-pst-fnct}
\int f'(x)d\mu_{\omega_{m_1}}^\perp(x) > 0~~a.e~~\omega_{m_1}.
\end{equation}
Given the Remark \ref{trnsl-inv} (in the appendix), w.l.o.g we can assume that the random variable $\omega_{m_1}$ can take positive values on a non-zero measure set and also negative values on a non-zero measure set.  Therefore using (\ref{innr-pst-fnct}) we can claim that 
\begin{equation}
\label{nn0-rvrale}
\begin{split}
&\omega_{m_1}\int f'(x)d\mu_{\omega_{m_1}}^\perp(x)
 -\mathbb{E}\bigg( \omega_{m_1}\int f'(x)d\mu_{\omega_{m_1}}^\perp(x) \bigg)\neq 0.\\
 & \qquad \text{The above is non zero as a random variable.}
 \end{split}
\end{equation}
Now the boundedness \big(it is discussed just below the (\ref{lim-exist})\big) of the r.h.s of (\ref{lim-exist}) will give the existence of the limit (as $k\to\infty$) of the square of the conditional expectation $\mathbb{E}\big(\Psi_{k,L_k,1,2,\cdots,d}(\omega) \big| \mathcal{D}_{1} \big)$ 
\begin{align}
\label{lim-sq=sq-lim}
&\lim_{k\to\infty}\bigg( \mathbb{E}\big(\Psi_{k,L_k,1,2,\cdots,d}(\omega) \big| \mathcal{D}_{1} \big) \bigg)^2\nonumber\\
&\qquad \qquad \qquad \qquad \qquad  =\bigg(\lim_{k\to\infty}\mathbb{E}\big(\Psi_{k,L_k,1,2,\cdots,d}(\omega) \big| \mathcal{D}_{1} \big)  \bigg)^2~a.e~\omega_{m_1}.
\end{align}
Since a non-zero random variable always has strictly positive second-order moment (about the origin),  therefore using (\ref{nn0-rvrale}), 
(\ref {lim-to-non-zero}) and (\ref{lim-exist}) in the above (\ref{lim-sq=sq-lim}) we get
\begin{align}
\label{ps-2-mnt}
\mathbb{E}\bigg[\lim_{k\to\infty}\bigg( \mathbb{E}\big(\Psi_{k,L_k,1,2,\cdots,d}(\omega) \big| \mathcal{D}_{1} \big) \bigg)^2 \bigg]>0.
\end{align}
Now using (\ref{sb-seq-lw-bd}),  (\ref{wr-vr-ag}) and Fatou's lemma we get
\begin{align}
\label{ag-ps-vr-pre}
\sigma^2_f&\ge \limsup_{k\to\infty}\mathbb{E}\bigg[\bigg( \mathbb{E}\big(\Psi_{k,L_k,1,2,\cdots,d}(\omega) \big| \mathcal{D}_{1} \big) \bigg)^2\bigg] \nonumber\\
&\ge \liminf_{k\to\infty}\mathbb{E}\bigg[\bigg( \mathbb{E}\big(\Psi_{k,L_k,1,2,\cdots,d}(\omega) \big| \mathcal{D}_{1} \big) \bigg)^2\bigg]\nonumber\\
&\ge \mathbb{E}\bigg[ \liminf_{k\to\infty}\bigg( \mathbb{E}\big(\Psi_{k,L_k,1,2,\cdots,d}(\omega) \big| \mathcal{D}_{1} \big) \bigg)^2 \bigg]\nonumber\\
&=\mathbb{E}\bigg[ \lim_{k\to\infty}\bigg( \mathbb{E}\big(\Psi_{k,L_k,1,2,\cdots,d}(\omega) \big| \mathcal{D}_{1} \big) \bigg)^2 \bigg]
>0.
\end{align}
In the last line of the above we have used (\ref{lim-exist}), (\ref{lim-sq=sq-lim}) and (\ref{ps-2-mnt}).  
Hence the lemma.
\end{proof}
\noindent Let $f\in C^1_P (\mathbb{R})$, and it is a monotone (strictly) function on an open interval $I$ which contains the spectrum $\sigma(H^\omega)~a.e ~\omega$. Now, we want to prove the strict positivity of the limiting variance $\sigma^2_f$ when the single site distribution (SSD) $\mu$ has non-compact support but satisfies the moment's condition $(\ref{mnts})$.  For that we define the measure $\bar{\nu}_n$ as
\begin{equation}
\label{p=1-mes-sm}
\bar{\nu}_n\big(\cdot\big)=\int_0^1\big[\mathbb{E}\big(\omega^2_n~\big\langle \delta_n, E_{H^\omega} \big(\cdot \big) \delta_n  \big\rangle\big|_{(\omega_n\to u\omega_n)}\big)\big]du,~~n\in\mathbb{Z}^d.
\end{equation}
Now from (\ref{p=1-mes-sm}),  (\ref{full-ldosm}), (\ref{fl-re-ldosm}),  (\ref{mes-r-at-n}) and the Corollary \ref{prv-meas-r-sm} (in appendix) it is clear that 
\begin{equation}
\label{tmny-eq-meas}
\bar{\nu}_n(\cdot)=\bar{\nu}_{p,n}(\cdot)=\bar{\nu}_p(\cdot)=\bar{\nu}(\cdot)~\forall~n\in\mathbb{Z}^d~~\text{and}~~p=1.
\end{equation}
\begin{lem}
\label{ps-sd-ncmpt}
Let the single site distribution (SSD) $\mu$ do not have compact support but satisfy the moment's condition (\ref{mnts}).  Assume $f\in C^1_P(\mathbb{R})$ is a strictly monotone function on an open interval $I$ (deterministic) such that $\sigma(H^\omega)\subseteq I~a.e~\omega$,  then the limiting variance $\sigma^2_f$ is positive (strictly).
\end{lem}
\begin{proof}
Since $f\in C^1_P(\mathbb{R})$ therefore the Lemma \ref{weak-con} for $p=1$ will give $f'\in L^2(\bar{\nu})$.  Now using (\ref{tmny-eq-meas}) and the Corollary \ref{pol-dens} we have a sequence of polynomials $\{P_k\}_{k=1}^\infty$ such that
\begin{equation}
\label{cn-in-l2}
  \|f'-P_k \|_{L^2(\bar{\nu}_n)}=\|f'-P_k \|_{L^2(\bar{\nu})}\to\ 0~~\text{as}~~k\to\infty~~\forall~n\in\mathbb{Z}^d.
\end{equation}
Lets denote $m_1=(1,1,\cdots,1)\in\mathbb{Z}^d$ and define the conditional expectations $Y_{m_1, k}$ and $Y_{f, m_1}$ as 
\begin{align}
\label{rv-cn-exp}
\begin{split}
& Y_{m_1, k}(\omega):=\mathbb{E}\bigg(\int_0^1\bigg(\omega_{m_1}\big\langle \delta_{m_1}, P_k\big( H^\omega \big)\delta_{m_1}  \big\rangle\big|_{(\omega_{m_1}\to u\omega_{m_1})}\bigg)du\bigg|\omega_{m_1}\bigg)\\
& Y_{m_1,f}(\omega):=\mathbb{E}\bigg(\int_0^1\bigg(\omega_{m_1}\big\langle \delta_{m_1}, f'\big( H^\omega \big)\delta_{m_1}  \big\rangle\big|_{(\omega_{m_1}\to u\omega_{m_1})}\bigg)du\bigg|\omega_{m_1}\bigg).
\end{split}
\end{align}
Given the Remark \ref{rs-fr-chg-int}, we can use Fubini's theorem together with total law of probability and Jensen's inequality for condition expectation to write
\begin{align}
\label{vr-est-for-ae}
&\mathbb{E}\bigg( Y_{m_1, k}-Y_{m_1, f} \bigg)^2\nonumber\\
&\leq \mathbb{E}\bigg( \mathbb{E}\bigg(\int_0^1\bigg(\omega_{m_1}\big\langle \delta_{m_1}, (P_k-f')\big( H^\omega \big)\delta_{m_1}  \big\rangle\big|_{(\omega_{m_1}\to u\omega_{m_1})}\bigg)^2du\bigg|\omega_{m_1}\bigg) \bigg)\nonumber\\
&=\mathbb{E}\bigg(\int_0^1\bigg(\omega_{m_1}\big\langle \delta_{m_1}, (P_k-f')\big( H^\omega \big)\delta_{m_1}  \big\rangle\big|_{(\omega_{m_1}\to u\omega_{m_1})}\bigg)^2du\bigg)\nonumber\\
&=\int_0^1\mathbb{E}\bigg(\omega_{m_1}\big\langle \delta_{m_1}, (P_k-f')\big( H^\omega \big)\delta_{m_1}  \big\rangle\big|_{(\omega_{m_1}\to u\omega_{m_1})}\bigg)^2du\nonumber\\
&\leq \int_0^1\mathbb{E}\bigg(\omega^2_{m_1}\big\langle \delta_{m_1}, \big|P_k-f'\big|^2\big( H^\omega \big)\delta_{m_1}  \big\rangle\big|_{(\omega_{m_1}\to u\omega_{m_1})}\bigg)du\nonumber\\
&=\int \big|P_k-f'\big|^2d\bar{\nu}_{m_1}(x),~~~~~~m_1=(1,1,\cdots,1)\in\mathbb{Z}^d
\end{align}
In the fourth inequality above, we have used the fact that for any self-adjoint operator $A$ on a Hilbert space $\mathcal{H}$ it is always true that $\langle \psi,  A\psi\rangle^2\leq \| \psi \|^2 \langle \psi,  A^2\psi\rangle~\forall~\psi\in\mathcal{H}$.   The definition of the finite measure $\bar{\nu}_{m_1}$ is given in (\ref{p=1-mes-sm}).\\
The convergence in (\ref{cn-in-l2}) give the convergence of $Y_{m_1, k}$ to $Y_{m_1, f}$ in the second order mean, i.e
\begin{equation}
\label{2nd-mean-cng}
\lim_{k\to\infty}\mathbb{E}\bigg( Y_{m_1, k}-Y_{m_1, f} \bigg)^2=0.
\end{equation}
An application of Markov inequality together with the above convergence (in second order mean) will give the convergence of $Y_{m_1, k}$ to $Y_{m_1, f}$ in probability.   Since the convergence in probability implies almost sure convergence through a subsequence, therefore we get
\begin{equation}
\label{cn-almst-thr-sub}
Y_{m_1, k_\ell}(\omega)\xrightarrow{\ell\to\infty} Y_{m_1,f}(\omega)~~a.e~\omega,~~\text{for some subsequence}~\{k_\ell\}~\text{of}~\{k\}.
\end{equation}
Using the definition (\ref{rv-cn-exp}), the above convergence can be re-written as the almost sure convergence of the conditional expectations, namely
 \begin{align}
 \label{pt-wse-cn-al}
 &\lim_{\ell\to\infty}\mathbb{E}\bigg(\int_0^1\bigg(\omega_{m_1}\big\langle \delta_{m_1}, P_{k_\ell}\big( H^\omega \big)\delta_{m_1}  \big\rangle\big|_{(\omega_{m_1}\to u\omega_{m_1})}\bigg)du\bigg|\omega_{m_1}\bigg)\nonumber\\
 &\qquad=\mathbb{E}\bigg(\int_0^1\bigg(\omega_{m_1}\big\langle \delta_{m_1}, f'\big( H^\omega \big)\delta_{m_1}  \big\rangle\big|_{(\omega_{m_1}\to u\omega_{m_1})}\bigg)du\bigg|\omega_{m_1}\bigg)~~a.e~\omega.
 \end{align}
 Again using the Jensen's inequality and (\ref{2nd-mean-cng}) we have
 \begin{align}
 \label{cn-in-mn}
 \lim_{\ell\to\infty}\bigg( \mathbb{E}\big(\big| Y_{m_1, k_\ell}-Y_{m_1, f}  \big|\big)\bigg)^2\leq \lim_{\ell\to\infty}\mathbb{E}\bigg( Y_{m_1, k_\ell}-Y_{m_1, f} \bigg)^2=0.
 \end{align}
 From above we get $\mathbb{E}\big(Y_{m_1,k_\ell}  \big)\to\mathbb{E}\big(Y_{m_1,f}  \big)$ as $\ell\to\infty$. 
 Now using the definition (\ref{rv-cn-exp}) and the total law of expectation we write
 \begin{align}
 \label{cn-in-mn-tl}
 &\lim_{\ell\to\infty}\mathbb{E}\bigg(\int_0^1\bigg(\omega_{m_1}\big\langle \delta_{m_1}, P_{k_\ell}\big( H^\omega \big)\delta_{m_1}  \big\rangle\big|_{(\omega_{m_1}\to u\omega_{m_1})}\bigg)du\bigg)\nonumber\\
 &\qquad=\mathbb{E}\bigg(\int_0^1\bigg(\omega_{m_1}\big\langle \delta_{m_1}, f'\big( H^\omega \big)\delta_{m_1}  \big\rangle\big|_{(\omega_{m_1}\to u\omega_{m_1})}\bigg)du\bigg).
 \end{align}
 Now define the polynomial $Q_{k_\ell}$ such that $Q'_{k_\ell}=P_{k_\ell}$ and define the trace of the operator $Q_{k_\ell}(H^\omega_L)$ as $\Psi_{k_\ell,L}(\omega)=Tr\big(Q_{k_\ell}(H^\omega_L)  \big)$,  here $\omega=(\omega_n)_{\in\mathbb{Z}^d}$.\\~\\
 Since we have from (\ref{cn-in-l2}) that $\|P_{k_\ell}-f' \|_{L^2(\bar{\nu})}\to 0$ as $\ell \to\infty$ therefore the same argument which have used to obtain (\ref{vr-f-lw}) will give
 \begin{equation}
 \label{lwbdd-unbdd}
 \sigma^2_f=\lim_{\ell\to\infty}\sigma^2_{Q_{k_\ell}}\ge \limsup_{\ell\to\infty}\limsup_{L\to\infty}
\mathbb{E}\big(\Psi^2_{k_\ell,L,1,2,\cdots,d}(\omega)  \big).
 \end{equation}
 Let the  function $h:\mathbb{N}\to\mathbb{N}$ defined as $h(k)=$ the  degree of the polynomial $P_k$.  W.l.o.g, we can also assume $h$ is a non-decreasing function on $\mathbb{N}$.  In view of the Remarks \ref{db-ge-arbitary} (in appendix) the above (\ref{lwbdd-unbdd}) can be re-written as
  \begin{equation}
 \label{lwbdd-unbdd-1}
 \sigma^2_f\ge \limsup_{\ell\to\infty}
\mathbb{E}\big(\Psi^2_{k_\ell,L_{k_\ell},1,2,\cdots,d}(\omega)  \big),~~L_{k_\ell}>\big(h(k_\ell)\big)^2.
 \end{equation}
 Now the same methods used in proving the Lemma \ref{pos-vr-mn-fn} will give our result instead of the entire sequence $\{k\}$ we have to work with its subsequence $\{k_\ell\}$.  Also we will use (\ref{pt-wse-cn-al}) and (\ref{cn-in-mn-tl}) in the step (\ref{der-mrt-estm}) to obtain the limit (\ref{lim-exist})\big(as given in the proof of the Lemma \ref{pos-vr-mn-fn}\big) and to show the limit in (\ref{lim-exist}) is finite a.e $\omega$ we will use the following statement
 \begin{align}
\bigg| \omega_{m_1}\int_0^1\mathbb{E}\bigg( \big\langle \delta_{m_1}, f'\big( H^\omega \big)\delta_{m_1}  \big\rangle\big|_{(\omega_{m_1}\to u\omega_{m_1})}\bigg|\omega_{m_1}\bigg)du\bigg|<\infty~~a.e~\omega.
 \end{align}
 The above will follow from the Remark \ref{rs-fr-chg-int}, the total law of expectation and the fact that the integrable (absolutely) function is finite almost surely.

\end{proof}
\noindent Now, to prove the CLT  (\ref{m-clt-dif-f}), we will only need to verify all the three conditions of the Theorem \ref{app-con-thm}.
\\~\\
{\bf Proof of Theorem \ref{clt-for-dif}:} Let assume $\sigma^2_f>0$ and also we have (\ref{prim-var-con}) therefore w.l.o.g we can assume $\sigma^2_k$ is positive for all $k$.  Now using the CLT (\ref{m-clt}) and the definition (\ref{prim-rv}) we write
\begin{equation}
\label{con1}
X_{Q_k,L}\xrightarrow[L\to\infty]{in~distribution} \mathcal{N}\big(0, \sigma^2_k  \big).
\end{equation}
We have also proved in (\ref{prim-var-con}) that
\begin{equation}
\label{con2}
\lim_{k\to\infty}\sigma^2_k=\sigma^2_f.
\end{equation}
And finally using Markov inequality,  (\ref{final-inq-imp}) and (\ref{dif-ran-der}), we write 
\begin{align}
\label{con3}
\mathbb{P}\bigg(\big| X_{f,L}-X_{Q_k,L} \big| \ge \delta \bigg)&\leq \frac{1}{\delta^2}\mathbb{E}\bigg( \big| X_{f,L}-X_{Q_k,L} \big|^2 \bigg),~~~\delta>0\nonumber\\
&=\frac{1}{\delta^2}\mathbb{E}\bigg( \big| X_{(f-Q_k),L} \big|^2 \bigg)\nonumber\\
&\leq \frac{\sqrt{8}}{\delta^2}\int \big| f'-P_k \big|^2d\bar{\nu}_L,~~~Q_k'=P_k.
\end{align}
Take $\limsup$ (w.r.t $L$) both sides of the above and use the Lemma \ref{weak-con} for $p=1$ to get
\begin{equation}
\label{con4}
\limsup_{L\to\infty}\mathbb{P}\bigg(\big| X_{f,L}-X_{Q_k,L} \big| \ge \delta \bigg)\leq \frac{\sqrt{8}}{\delta^2}\int \big| f'-P_k \big|^2d\bar{\nu}.
\end{equation}
In (\ref{poly-apprx-spect}) it is given that $\| f'-P_k \|_{L^2(\bar{\nu})}\to 0$ as $k\to\infty$,  so in (\ref{con1}), (\ref{con2}) and (\ref{con4}) we verified all the conditions of the Theorem \ref{app-con-thm}.\\~\\
For $f\in C^1_P(\mathbb{R})$, the finiteness of the limiting variance $\sigma^2_f$ is given in the Proposition \ref{fin-diff}.  
For strictly monotone function (on an open non-random interval $I$ contains $\sigma(H^\omega)$) $f\in C^1_P(\mathbb{R})$,  the positivity (strictly) of the limiting variance $\sigma^2_f$ has already been proved in the Lemma \ref{pos-vr-mn-fn}, for compactly supported $\mu$ (SSD) and in the Lemma \ref{ps-sd-ncmpt}, for non-compactly supported $\mu$ (SSD).  Hence, we got the proof of our result.
\qed
\appendix
\section{Appendix}
We proved some results in this appendix, which are used in the main part of the paper.  Some of these results might be known in the literature, but we proved it in the form we need.\\
First, we obtain a formula for the derivative of the trace of a matrix w.r.t its diagonal elements.
Let $T$ be a self-adjoint operator on a finite-dimensional Hilbert space $\mathcal{H}$ and denote $\{\delta_n\}_{n=1}^m$ be a orthonormal basis of $\mathcal{H}$, $m<\infty$.  Now define the rank one perturbation $T_\lambda=T+\lambda \mathcal{P}_n,~\lambda\in\mathbb{R}$. 
Here $\mathcal{P}_n$ denote the projection onto the subspace generated by the single vector 
$\delta_n$, i.e $\mathcal{P}_n\varphi=\langle\varphi, \delta_n\rangle \delta_n~\forall~\varphi\in\mathcal{H}$.
\begin{lem} 
Let $T$,  $T_\lambda$ and $\mathcal{H}$ as defined above then for any $f\in C^1(\mathbb{R})$, we have
\label{der-tr}
\begin{equation}
\label{tr-fr}
\frac{\partial}{\partial \lambda}\big(Tr\big( f(T_\lambda) \big)  \big)=\langle f'(T_\lambda)\delta_n, \delta_n\rangle.
\end{equation}
 \end{lem}
 \begin{proof}
 Let $\{E_k\}_{k=1}^m$ be the eigenvalues and $\{\psi_k\}_{k=1}^m$ are the corresponding eigenfunctions of $T_\lambda$ then by the spectral theorem we write
 \begin{equation}
 \label{sp-decom}
 f(T_\lambda)=\sum_{k=1}^m f(E_k) \mathcal{Q}_{_k},~~\text{here}~~\mathcal{Q}_{_k}\varphi=\langle\varphi, \psi_k\rangle \psi_k~\forall~\varphi\in\mathcal{H}.
 \end{equation}
The trace of the operator $f(T_\lambda)$ and its derivative is given by
 \begin{equation}
 \label{tr-T}
 Tr\big( f(T_\lambda) \big)=\sum_{k=1}^m f(E_k)~~and~~
 \frac{\partial}{\partial \lambda}\big(Tr\big( f(T_\lambda) \big)  \big)=\sum_{k=1}^m f'(E_k)\frac{\partial E_k}{\partial \lambda}.
 \end{equation}
 Now the derivative of the eigenvalue $E_k$ w.r.t the parameter $\lambda$ is given by Hellmann-Feynman theorem \cite[equation (2.4)]{IM}
 \begin{equation}
 \label{hell-fey}
 \frac{\partial E_k}{\partial \lambda}=\big|\langle \psi_k, \delta_n\rangle  \big|^2,~~\text{for}~k=1,2,\cdots, m.
 \end{equation}
 So the derivative of the trace w.r.t $\lambda$ can be written as
 \begin{equation}
 \label{deri-trace}
  \frac{\partial}{\partial \lambda}\big(Tr\big( f(T_\lambda) \big)  \big)=\sum_{k=1}^m f'(E_k)\big|\langle \psi_k, \delta_n\rangle  \big|^2.
\end{equation}
For $f\in C^1(\mathbb{R})$,  again by spectral theorem we write 
\begin{equation}
\label{deri-spec}
f'(T_\lambda)=\sum_{k=1}^m f'(E_k) \mathcal{Q}_{_k}~~\text{and}~~
\langle f'(T_\lambda)\delta_n, \delta_n\rangle=\sum_{k=1}^m f'(E_k)\big|\langle \psi_k, \delta_n\rangle  \big|^2.
\end{equation}
Now, the lemma will follow from (\ref{deri-trace}) and (\ref{deri-spec}).
 \end{proof}
 \begin{rem}
The above lemma is also true in infinite-dimensional Hilbert space $\mathcal{H}$ as long as the operator $T$ is compact and the trace of $f(T_\lambda)$ is finite. 
 \end{rem}
 \noindent Now we will prove a inequality involving limsup of a double sequence.
 \begin{prop}
 \label{doule-seq}
 Let $\{x_{n,m}\}_{n,m}$ is a double sequence of positive real numbers such that $\displaystyle\limsup_{n\to\infty}\displaystyle\limsup_{m\to\infty}x_{n,m}<\infty$ then there exist a strictly increasing subsequence $\{m_n\}_{n}$ of $\{m\}_m$ having property $m_n>n^2$ such that
 \begin{equation}
 \label{limsp-d-seq}
 \limsup_{n\to\infty}x_{n,m_n}\leq \limsup_{n\to\infty}\limsup_{m\to\infty}x_{n,m}.
 \end{equation}
 \end{prop}
 \begin{proof}
 Since we have  $0\leq \displaystyle\limsup_{n\to\infty}\bigg(\displaystyle\limsup_{m\to\infty}x_{n,m}\bigg)<\infty$ therefore using the definition of the $\limsup$ (largest limit point) of a sequence we get that  
 \begin{equation}
 \label{chose-incr}
 \begin{split}
& 0\leq \limsup_{m\to\infty}x_{n,m} <\infty~~\text{for each}~n\in\mathbb{N}~\text{and also }\\
 & \big|x_{n,m}-\limsup_{m\to\infty}x_{n,m}\big|<\frac{1}{n}~\text {holds for infinitely many } m,~\text{for each } n\in\mathbb{N}.
\end{split}
 \end{equation}
 Chose $1<m_1\in \mathbb{N}$ such that $\big|x_{1,m_1}-\displaystyle \limsup_{m\to\infty}x_{1,m}\big|<1$ and in view of (\ref{chose-incr}) we can again choose $m_2>\max\{2^2, m_1\}$ such that $\big|x_{2,m_1}-\displaystyle \limsup_{m\to\infty}x_{2,m}\big|<\frac{1}{2}$ therefore by induction we always have
 \begin{equation}
 \label{ind-cho}
 \big|x_{n,m_n}- \limsup_{m\to\infty}x_{n,m}\big|<\frac{1}{n},~~\text{here} ~m_n>\max\{n^2, m_{n-1}\}~\text{and}~n\in\mathbb{N}.
 \end{equation}
 Now we write
 \begin{align}
 \label{cl-db-lmsp}
 0\leq x_{n,m_n}=\big(x_{n,m_n}-\limsup_{m\to\infty}x_{n,m}\big)+\limsup_{m\to\infty}x_{n,m}.
 \end{align}
 Now (\ref{limsp-d-seq}) is immediate once we use (\ref{ind-cho}) in the above (\ref{cl-db-lmsp}).
  \end{proof}
  \begin{rem}
  \label{db-ge-arbitary}
Let $h:\mathbb{N}\to\mathbb{N}$ be a function such that $h(n)\leq h(n+1),~\forall~n\in\mathbb{N}$ then using the same arguments used in the above proposition, we can also choose $m_n>(h(n))^2$ in (\ref{limsp-d-seq}).
\end{rem}
\noindent We want to note that if we translate each random variable $\omega_n$ by a fixed deterministic constant structurally, there would not be any changes in all the results we have proved above.
\begin{rem}
\label{trnsl-inv}
 Let $b\in \mathbb{R}$ be a fixed non-random constant.  Consider the collection of i.i.d random variables $\{\tilde{\omega}_n\}_{n\in\mathbb{Z}^d}$, here $\tilde{\omega}_n=\omega_n-b$ and define the random Schr\"{o}dinger operator $H^{\tilde{\omega}}$ on $\ell^2(\mathbb{Z}^d)$ as
 $$H^{\tilde{\omega}}=\big(\Delta+b \mathbb{I}\big)+V^{\tilde{\omega}},~~\text{here}~~ \big(V^{\tilde{\omega}}u\big)(n)=\tilde{\omega}_nu(n),~u\in\ell^2(\mathbb{Z}^d).$$
 In the above $\mathbb{I}$ is the identity operator on $\ell^2(\mathbb{Z}^d)$.  The discrete Laplacian $\Delta$ and the collection of i.i.d random variables $\{\omega_n\}_{n\in\mathbb{Z}^d}$ are the same as it is in (\ref{model}).\\
Now it is immediate that for any realization of $\{\omega_n\}_{n\in\mathbb{Z}^d}$ $(\text{or}~\{\tilde{\omega}_n\}_{n\in\mathbb{Z}^d})$ we always have $H^\omega=H^{\tilde{\omega}}$.
\end{rem}
\noindent We will prove that the covariance of two successive martingale (Doob) differences is always zero. Let $P(X)$ be a real polynomial function depends on a collection of independent random variables $X=(X_n)_{n\in\Lambda}$ with $|\Lambda|<\infty$ and each $X_n$ is defined on the same probability space.   Let $\{A_k\}_{k=1}^m$ is a collection of subset of $\Lambda$ such that $A_k\subset A_{k+1}$ and $A_m=\Lambda$.   Denote $\mathcal{F}_k:=\sigma\big(X_n: n\in A_k  \big)$, the $\sigma$-algebra generated by the collection of independent random variables $(X_n)_{n\in A_k}$ and it is immediate that this collection of $\sigma$-algebra $\{\mathcal{F}_k \}_{k=1}^m$ is a filtration.
 \begin{prop}
 \label{or-db-dif-mart}
 Consider $P(X)$ as multi-variable polynomial in $X=\big\{X_n  \big\}_{n\in\Lambda}$, satisfy the condition $\mathbb{E}\big(\big| P(X) \big| \big)<\infty$ and $\{\mathcal{F}_k\}_{k=1}^m$ are the $\sigma$-algebras as described above then the collection of conditional expectations
 $\big\{\mathbb{E}\big( P(X)\big| \mathcal{F}_k\big) \big\}_{k=1}^m$ form a martingale (Doob) and the variance of the random variable $P(X)$ can be given by the formula
 \begin{equation}
 \label{vr-db-martgl}
 \mathbb{E}\bigg(P(X)-\mathbb{E}\big( P(X) \big)  \bigg)^2=\sum_{k=1}^m \mathbb{E}\bigg( \mathbb{E}\big( P(X)\big| \mathcal{F}_k\big) -\mathbb{E}\big( P(X)\big| \mathcal{F}_{k-1}\big)\bigg)^2.
 \end{equation}
 In the above, we denote $\mathcal{F}_0$ as the trivial $\sigma$-algebra (consists of empty and total space).
 \end{prop}
\begin{proof}
It is immediate that the collection of random variables $\big\{\mathbb{E}\big( P(X)\big) \big\}_{k=1}^m$ form a martingale also it is true that $\mathbb{E}\bigg(\mathbb{E}\big( P(X)\big| \mathcal{F}_j\big)\bigg|\mathcal{F}_i  \bigg)=\mathbb{E}\big( P(X)\big| \mathcal{F}_i\big)$ for $i<j$.  Now the difference between $f(X)$ and $\mathbb{E}\big( f(X)\big)$ can be written as
\begin{align}
\label{sum-diffrnce}
P(X)-\mathbb{E}\big( P(X)\big)&=\sum_{k=1}^m \bigg(\mathbb{E}\big( P(X)\big| \mathcal{F}_k\big) -\mathbb{E}\big( P(X)\big| \mathcal{F}_k\big)\bigg)\nonumber\\
&=\sum_{k=1}^m \big(Y_k-Y_{k-1}\big),~~Y_k=\mathbb{E}\big( P(X)\big| \mathcal{F}_k\big).
\end{align}
It is easy to observe that $\mathbb{E}\big(Y_k-Y_{k-1}  \big)=0$,  law of total expectation.  Now we will show that the covariance between the two random variables $\big( Y_i-Y_{i-1} \big)$ and $\big( Y_i-Y_{j-1} \big)$ is always zero for $i< j$. Again, using the law of total expectation we have
\begin{align}
\label{lte-use}
\mathbb{E}\bigg(\big(Y_i-Y_{i-1}  \big)\big(Y_j-Y_{j-1}  \big)  \bigg)&=\mathbb{E}\bigg(\mathbb{E}\big(\big(Y_i-Y_{i-1}  \big)\big(Y_j-Y_{j-1}  \big)\big|\mathcal{F}_i  \big)\bigg)\nonumber\\
& =\mathbb{E}\bigg(\big(Y_i-Y_{i-1}  \big)\mathbb{E}\big(\big(Y_j-Y_{j-1}  \big)\big|\mathcal{F}_i  \big)\bigg)\nonumber\\
&=\mathbb{E}\bigg(\big(Y_i-Y_{i-1}  \big)\big(Y_i-Y_{i}  \big)\bigg)\nonumber\\
&=0.
\end{align}
In the second equality above we used the fact that $\mathbb{E}\big(Z_1 Z_2\big| \mathcal{F}  \big)=Z_1 \mathbb{E}\big(Z_2\big| \mathcal{F}  \big)$ whenever $Z_1$ is $\mathcal{F}$-measurable random variable.\\
Now (\ref{vr-db-martgl}) will follow from (\ref{sum-diffrnce}) and (\ref{lte-use}). Hence the proposition.
\end{proof}
\noindent Let $A$ and $B$ be two subsets of the indexing set $\Lambda$, as in the above.  Denote $\mathcal{F}_A=\sigma\big(X_n: n\in A  \big)$,  the $\sigma$-algebra generated by the collection of independent random variables $\big\{ X_n \big\}_{n\in A}$ and the same is true for $\mathcal{F}_B$.   Define $\mathcal{F}_A\mathcal{F}_B=\sigma\bigg( X_n:n\in A\cap B \bigg)$, the $\sigma$-algebra generated by the collection of independent random variables $\big\{ X_n \big\}_{n\in A\cap B}$.
Now, we have the following identity concerning conditional expectations.
\begin{prop}
\label{con-ex-id}
Let $P(X)$ and the two $\sigma$-algebras $\mathcal{F}_A$, $\mathcal{F}_B$ as defined above then we always have
\begin{equation}
\label{idty-ex-con}
\mathbb{E}\bigg(\mathbb{E}\big( P(X)\big| \mathcal{F}_A \big)\bigg|\mathcal{F}_B\bigg)=\mathbb{E}\big( P(X)\big| \mathcal{F}_A \mathcal{F}_B\big),
\end{equation}
here $X=(X_n)_{n\in\Lambda},~|\Lambda|<\infty$ is a collection of independent random variables and both $A$, $B$ are subsets of $\Lambda$.
\end{prop}
\begin{proof}
The proof will quickly follow from the independence of $(X_n)_{n\in\Lambda}$ and the property of conditional expectation, namely
$\mathbb{E}\big(Z_1 Z_2\big| \mathcal{F}  \big)=Z_1 \mathbb{E}\big(Z_2\big| \mathcal{F}  \big)$, when $Z_1$ is a $\mathcal{F-}\text{measurable}$ random variable.
\end{proof}
\noindent To estimate the moments of the finite measure $\bar{\nu}(\cdot)$ \big(as in (\ref{ldosm})\big) we need to find bounds of the expected value of the random variable $\big\langle \delta_0,   (H^\omega)^k\delta_0\big\rangle$ $\forall$ $k\in\mathbb{N}$.   To do that we are going to write $\big\langle \delta_0,   (H^\omega)^k\delta_0\big\rangle$ as a multi-variable polynomial in $\{\omega_n\}_{n\in\mathbb{Z}^d}$, which is describe below.
\begin{rem}
\label{poly-op} Let $P(x)=\displaystyle \sum_{k=0}^pa_k ~x^k$ be a polynomial of degree $p$.  Then, the spectral theorem of the self-adjoint operator will give
\begin{equation}
\label{spt}
\langle \delta_n, P(H^\omega)\delta_n\rangle=\sum_{k=0}^pa_k\langle \delta_n, (H^\omega)^k\delta_n\rangle~~\forall~~n\in\mathbb{Z}^d.
\end{equation}
Using definition (\ref{model}),  it is also possible to write each monomial (in operator) $ \langle \delta_n, (H^\omega)^k\delta_n\rangle,~k\in\mathbb{N}$ in the form as 
\begin{equation}
\label{mono}
\langle \delta_n, (H^\omega)^k\delta_n\rangle=
\sum_{\substack{n_i\in\mathbb{Z}^d\\|n_i-n|\leq k\\i=1,2,\cdots,k}}
\sum_{\substack{j_i\in\mathbb{N}\cup\{0\},~j_i\leq j_{i+1}\\0\leq j_1+\cdots+j_k\leq k\\i=1,2,\cdots,k}}C^{j_1,j_2,\cdots j_k}_{n_1,n_2,\cdots,n_k}~~\omega_{n_1}^{j_1}\omega_{n_2}^{j_2}\cdots\omega_{n_k}^{j_k},
\end{equation}
here $C^{j_1,j_2,\cdots j_k}_{n_1,n_2,\cdots,n_k}\ge 0$ are the non-negative constants depend on the multi-indices $\{n_i\}_{i=1}^k$ and $\{j_i\}_{i=1}^d$ but  they are translation invariant on $\mathbb{Z}^d$, i.e for each $m\in\mathbb{Z}^d$ we have $C^{j_1,j_2,\cdots j_k}_{n_1,n_2,\cdots,n_k}=C^{j_1,j_2,\cdots j_k}_{n_1-m,n_2-m,\cdots,n_k-m}$.\\~\\
For the finite-dimensional approximation $H^\omega_L$ as in (\ref{restric}) the expression of $ \langle \delta_n, (H^\omega_L)^k\delta_n\rangle$, $n\in\Lambda_L$ can be given as
\begin{equation}
\label{mono-fn}
\langle \delta_n, (H^\omega_L)^k\delta_n\rangle=
\sum_{\substack{n_i\in\Lambda_L\\|n_i-n|\leq k\\i=1,2,\cdots,k}}
\sum_{\substack{j_i\in\mathbb{N}\cup\{0\},~j_i\leq j_{i+1}\\0\leq j_1+\cdots+j_k\leq k\\i=1,2,\cdots,k}}C^{j_1,j_2,\cdots j_k}_{n_1,n_2,\cdots,n_k,  L}~~\omega_{n_1}^{j_1}\omega_{n_2}^{j_2}\cdots\omega_{n_k}^{j_k},
\end{equation}
here $C^{j_1,j_2,\cdots j_k}_{n_1,n_2,\cdots,n_k, L}$ are the non negative constants satisfy the inequality 
$$0\leq C^{j_1,j_2,\cdots j_k}_{n_1,n_2,\cdots,n_k, L}\leq C^{j_1,j_2,\cdots j_k}_{n_1,n_2,\cdots,n_k}.$$
\end{rem}
 \noindent For each $p\in\mathbb{N}\cup \{0\}$ we define the finite measure $\bar{\nu}_{p,L}(\cdot)$ on $\mathbb{R}$ as
 \begin{align}
 \label{re-ldosm}
 \bar{\nu}_{p,L}\big(\cdot\big)=\frac{1}{|\Lambda_L|}\sum_{n\in\Lambda_L}\int_0^1\big[\mathbb{E}\big(\omega^{2p}_n~\big\langle \delta_n, E_{H^\omega_L} \big(\cdot \big) \delta_n  \big\rangle\big|_{(\omega_n\to u\omega_n)}\big)\big]du.
 \end{align}
 The corresponding measure associated with full operator $H^\omega$  is defined by
 \begin{align}
 \label{fl-re-ldosm}
 \bar{\nu}_{p}\big(\cdot\big)=\int_0^1\big[\mathbb{E}\big(\omega^{2p}_0~\big\langle \delta_0, E_{H^\omega} \big(\cdot \big) \delta_n \big\rangle\big|_{(\omega_0\to u\omega_0)}\big)\big]du.
 \end{align}
 The total mass of the above measures is finite,  i.e $\bar{\nu}_{p,L}(\mathbb{R})=\bar{\nu}_p(\mathbb{R})=\mathbb{E}\big(\omega^{2p}_0 \big)$.\\~\\
\noindent To show the variance $\sigma^2_f$ is finite for every $f$ in $C^1_p(\mathbb{R})$ (see Definition \ref{def}) we need to prove that $f'\in L^2(\bar{\nu})~\forall~f\in C^1_p(\mathbb{R})$,  $\bar{\nu}$ is the same measure as $\bar{\nu}_{p}$ for $p=1$ .  It will follow if we show that each moment of $\bar{\nu}$ is finite. 
\begin{prop}
\label{mnt-det}
Let the single site distribution (SSD) $\mu$ satisfy the moments estimation (\ref{mnts}) then the measure (finite) $\bar{\nu}_p$ is determined by its moments.
\end{prop}
\begin{proof}
Since the measure 
$\bar{\nu}_p(\cdot)=\int_0^1\big[\mathbb{E}\big(\omega_0^{2p}\big\langle \delta_0, E_{H^\omega}(\cdot)\delta_0\big\rangle\big|_{(\omega_0\to u\omega_0)}\big)\big]du$ therefore using the spectral theorem and (\ref{mono}) we write
\begin{align}
\label{mnt-dm}
\int x^k d\bar{\nu}_p(x)\nonumber& =\int_0^1\big[\mathbb{E}\big(\omega_0^{2p}\big\langle \delta_0, (H^\omega)^k\delta_0\big\rangle\big|_{(\omega_0\to u\omega_0)}\big)\big]du\nonumber\\
& =\sum_{\substack{n_i\in\mathbb{Z}^d\\|n_i|\leq k\\i=1,2,\cdots,k}}
\sum_{\substack{j_i\in\mathbb{N}\cup\{0\},~j_i\leq j_{i+1}\\0\leq j_1+\cdots+j_k\leq k\\i=1,2,\cdots,k}}
 \bigg[C^{j_1,j_2,\cdots j_k}_{n_1,n_2,\cdots,n_k}~\times\nonumber\\
&\qquad \qquad \bigg(\int_0^1 u^{m(j_1,j_2,\cdots,j_k)}du\bigg)~ \mathbb{E}\big(\omega_0^{2p}\omega_{n_1}^{j_1}\omega_{n_2}^{j_2}\cdots\omega_{n_k}^{j_k}\big)\bigg].
\end{align}
In the above we denote $m(j_1,j_2,\cdots,j_k)=j_i$ if $n_i=0$ for some $i=1,2,\cdots,k$ otherwise it is $0$.\\
Using the bound (\ref{mnts}) and the independence of $\{\omega_n\}_{n\in\mathbb{Z}^d}$ we estimate the expectation inside the above sum as
\begin{align}
\label{avg-mnt}
\begin{split}
\bigg| \mathbb{E}\big(\omega_0^{2p}\omega_{n_1}^{j_1}\omega_{n_2}^{j_2}\cdots\omega_{n_k}^{j_k}\big)\bigg|&\leq \prod_{i=1}^kCa^{\tilde{j}_i}\tilde{j}^{\tilde{j}_i}\leq C^k a^{k+2p} (k+2p)^{k+2p}, \\
\text{here we have } ~& 0\leq \tilde{j}_1+\tilde{j}_2+\cdots+\tilde{j}_k\leq k+2p.
\end{split}
\end{align}
Let $H=\Delta+V$ be the deterministic operator on $\ell^2(\mathbb{Z}^d)$ with the same discrete Laplacian $\Delta$ as defined in (\ref{model}) and here $V$ is the identity operator.  Now we can write the vector $H^k\delta_0$ as the finite linear combination of the basis elements $\{ \delta_m\}_{|m|\leq k}$
\begin{equation}
\label{lin-conb}
H^k\delta_0=\sum_{|m|\leq k} c_m \delta_m,~\text{where}~c_m\in\mathbb{N}\cup\{0\}~\text{and}~\sum_{|m|\leq k} c_m=(2d+1)^k.
\end{equation}
It is also true from above that
\begin{equation}
\label{inn-co}
\langle \delta_0,  H^k \delta_0\rangle=c_0~~\text{with}~~0\leq c_0\leq(2d+1)^k.
\end{equation}
From (\ref{mono}) the expression of $\langle \delta_0,  (H^\omega)^k \delta_0\rangle$ can be written as
\begin{equation}
\label{mono-ag}
\langle \delta_0, (H^\omega)^k\delta_0\rangle=
\sum_{\substack{n_i\in\mathbb{Z}^d\\|n_i|\leq k\\i=1,2,\cdots,k}}
\sum_{\substack{j_i\in\mathbb{N}\cup\{0\},~j_i\leq j_{i+1}\\0\leq j_1+\cdots+j_k\leq k\\i=1,2,\cdots,k}}C^{j_1,j_2,\cdots j_k}_{n_1,n_2,\cdots,n_k}~~
\omega_{n_1}^{j_1}\omega_{n_2}^{j_2}\cdots\omega_{n_k}^{j_k}.
\end{equation}
But it is immediate that $\langle \delta_0, (H^\omega)^k\delta_0\rangle=\langle \delta_0,  H^k \delta_0\rangle$ for
 $\omega_n=1~\forall~n\in\mathbb{Z}^d$ in  (\ref{model}),  the definition of $H^\omega$.  So using (\ref{inn-co}) and (\ref{mono-ag}) we have
\begin{equation}
 \label{est-sum-cnt}
 0\leq\sum_{\substack{n_i\in\mathbb{Z}^d\\|n_i|\leq k\\i=1,2,\cdots,k}}
\sum_{\substack{j_i\in\mathbb{N}\cup\{0\},~j_i\leq j_{i+1}\\0\leq j_1+\cdots+j_k\leq k\\i=1,2,\cdots,k}}C^{j_1,j_2,\cdots j_k}_{n_1,n_2,\cdots,n_k}\leq (2d+1)^k.
 \end{equation}
 We also have a simple estimation
 \begin{align}
 \label{smple-est}
 \int_0^1 u^m du=\frac{1}{m+1}\leq 1,~~m\in\mathbb{N}\cup\{0\}.
 \end{align}
Using (\ref{smple-est}), (\ref{est-sum-cnt}) and (\ref{avg-mnt}) in (\ref{mnt-dm}), we can give an estimation on the moments of the density of states measure (DOSm) $\nu$ as \begin{equation}
 \label{bd-mnt-dm}
 |m_k|\leq (2d+1)^kC^ka^{k+2p}(k+2p)^{k+2p},~~\text{where}~~m_k=\int x^k d\bar{\nu}_p(x).
 \end{equation}
Lets define the power series $\displaystyle\sum_k\frac{m_k}{k!}t^k$.  Now, its radius of convergence $r$ is given by 
\begin{equation}
r=\bigg( \limsup_k \bigg(\frac{|m_k|}{k!}\bigg)^{\frac{1}{k}}\bigg)^{-1}\ge \frac{1}{(2d+1)C a e}>0,
\end{equation}
here, we have used Stirling's approximation for the $k!$.  Now the moments determinacy of the finite measure $\bar{\nu}_p$ will follow from  \cite[Theorem 30.1]{PB}.
\end{proof}
\noindent For $p\in \mathbb{N}\cup\{0\}$,  also we can define similar measure as in (\ref{fl-re-ldosm}) associated with the vector $\delta_n$ as
\begin{equation}
\label{mes-r-at-n}
\bar{\nu}_{p,n}\big(\cdot\big)=\int_0^1\big[\mathbb{E}\big(\omega^{2p}_n~\big\langle \delta_n, E_{H^\omega} \big(\cdot \big) \delta_n  \big\rangle\big|_{(\omega_n\to u\omega_n)}\big)\big]du,~~n\in\mathbb{Z}^d.
\end{equation}
So from (\ref{mes-r-at-n}) and (\ref{fl-re-ldosm}) we have $\bar{\nu}_p(\cdot):=\bar{\nu}_{p,0}(\cdot)$.   Since $\{\omega_n\}_{n\in\mathbb{Z}^d}$ are i.i.d real random variables,  so it can be proved that $\bar{\nu}_{p,n}(\cdot)=\bar{\nu}_{p,0}(\cdot)=:\bar{\nu}_p(\cdot)~\forall~n\in\mathbb{Z}^d$.
\begin{cor}
\label{prv-meas-r-sm}
From the independent identical distribution of $\{\omega_n\}_{n\in\mathbb{Z}^d}$ and (\ref{mono}) it is easy to observe that for any $n\in\mathbb{Z}^d$ we always have
\begin{align}
\label{mnt-r-eq}
\int x^k\bar{\nu}_p(x):&=\int x^k\bar{\nu}_{p,0}(x),~~~~\text{here}~~k\in\mathbb{N}\cup\{0\}\nonumber\\
&=\int_0^1\big[\mathbb{E}\big(\omega^{2p}_0~\big\langle \delta_0, (H^\omega)^k 
 \delta_0  \big\rangle\big|_{(\omega_0\to u\omega_0)} \big)\big]du\nonumber\\
&=\int_0^1\big[\mathbb{E}\big(\omega^{2p}_n~\big\langle \delta_n, (H^\omega)^k 
 \delta_n  \big\rangle\big|_{(\omega_n\to u\omega_n)} \big)\big]du\nonumber\\
 &=\int x^k\bar{\nu}_{p,n}(x).
\end{align}
From definition (\ref{mes-r-at-n}) it is clear that both the measure $\bar{\nu}_{p,n}(\cdot)$ and $\bar{\nu}_p(\cdot)$ are finite measures.
Also, from the above (\ref{mnt-r-eq}), we know that all the moments of the two measures $\bar{\nu}_{p,n}(\cdot)$ and $\bar{\nu}_p(\cdot)$ are same.  Since the measure $\bar{\nu}_p(\cdot)$ is determined by its moments so we have the equality  $\bar{\nu}_{p,n}(\cdot)=\bar{\nu}_p(\cdot),~\forall~n\in\mathbb{Z}^d$.
\end{cor}
\begin{cor}
\label{pol-dens}
Since under the condition (\ref{mnts}), we know that the finite measure $\bar{\nu}_p$ is moments determinate, then the set of all polynomials is dense in $L^2(\bar{\nu}_p)$ by \cite[Corolarry 2.50]{DPA}.  
 \end{cor}
 \begin{cor}
 \label{mnt-det-dosm}
 The similar computation as we did in (\ref{bd-mnt-dm}) for the finite measure $\bar{\nu}_p(\cdot)$ will give the moments determinacy of the density of state measure (DOSm) $\nu(\cdot)=\mathbb{E}\big( \big\langle \delta_0, E_{H^\omega}(\cdot)\delta_0\big\rangle \big)$.
 \end{cor}
\noindent Now we will show the weak convergence of the sequence of finite measures $\{\bar{\nu}_{p,L}(\cdot)\}_L$ \big(as in (\ref{re-ldosm}) \big) to the finite measure $\bar{\nu}_p(\cdot)$.
\begin{lem}
\label{weak-con}
Let the single site distribution (SSD) $\mu$ satisfy the moment condition (\ref{mnts}) and $f\in C^1_P(\mathbb{R})$ then
$f'\in L^2(\bar{\nu}_p)\cap L^2({\bar{\nu}_{p,L}})$ and we also have
\begin{equation}
\label{cng-w}
\int \big|f'(x)\big|^2d\bar{\nu}_{p,L}(x)\xrightarrow{L\to\infty} \int \big|f'(x)\big|^2d\bar{\nu}_p(x),~~\forall~f\in C^1_P(\mathbb{R}).
\end{equation}
\end{lem}
\begin{proof}
Since $|f'(x)|\leq P(x)$ for some polynomial $P(x)$ for $f\in C^1_P(\mathbb{R})$ then from (\ref{bd-mnt-dm}) it will easily follow that $f'\in L^2(\bar{\nu}_p)$.  Now to prove $f'\in L^2(\bar{\nu}_{p,L})$ it is enough to show 
\begin{equation}
\label{mnt-fin-l}
\bigg|\int x^kd\bar{\nu}_{p,L}(x)\bigg|<\infty~~\forall~k\in\mathbb{N}~~\text{and}~~L\ge1.
\end{equation}
Using the definition (\ref{ldosm}) and the spectral theorem, we can write the moments of $\bar{\nu}_L$ as
\begin{equation}
\label{mnt-thr-spec}
\int x^kd\bar{\nu}_{p,L}(x)=\frac{1}{|\Lambda_L|}\sum_{n\in\Lambda_L}\int_0^1\big[\mathbb{E}\big(\omega_n^{2p}\big\langle \delta_n, (H^\omega_L)^k\delta_n\big\rangle\big|_{(\omega_n\to u\omega_n)}\big)\big]du.
\end{equation}
Now if we look at the expressions (\ref{mono}) and (\ref{mono-fn}) for $\big\langle \delta_n, (H^\omega)^k\delta_n \big\rangle$ and 
$\big\langle \delta_n, (H^\omega_L)^k\delta_n \big\rangle$, respectively, we can observe that both the expressions are the same except for the coefficients of
$\omega_{n_1}^{j_1}\omega_{n_2}^{j_2}\cdots\omega_{n_k}^{j_k}$. It is also true that (see (\ref{mono-fn})) the coefficients in the expression of $\big\langle \delta_n, (H^\omega_L)^k\delta_n \big\rangle$ are smaller than the that of 
$\big\langle \delta_n, (H^\omega)^k\delta_n \big\rangle$.  
Hence, using the exact same method as we did in (\ref{bd-mnt-dm}) and (\ref{mnt-dm}), we can actually show that
\begin{align}
\label{est-ldsm}
&\bigg|\int_0^1\big[\mathbb{E}\big(\omega_n^{2p}\big\langle \delta_n, (H^\omega_L)^k\delta_n\big\rangle\big|_{(\omega_n\to u\omega_n)}\big)\big]du\bigg|\nonumber\\
&\qquad \qquad \qquad ~\leq (2d+1)^kC^ka^{k+2p}(k+2p)^{k+2p}~~\forall~n\in\Lambda_L~\text{and}~L\ge 1.
\end{align}
Now it is clear from (\ref{mnt-thr-spec}) that
\begin{equation}
\label{fin-mnt-ldsm}
\bigg|\int x^kd\bar{\nu}_{p,L}(x)\bigg|\leq (2d+1)^kC^ka^{k+2p}(k+2p)^{k+2p}<\infty~~\forall~L\ge1.
\end{equation}
So we get $f'\in L^2(\bar{\nu}_p)\cap L^2(\bar{\nu}_{L,p})$ whenever $f\in C^1_P(\mathbb{R})$.\\~\\
We know the probability measure $\bar{\nu}_p(\cdot)$ is determined by its moments and $|f'(x)|\leq P(x)$ for some polynomial $P(x)$ for $f\in C^1_P(\mathbb{R})$. Therefore the convergence (\ref{cng-w}) is direct consequence of \cite[Lemma 2.1]{DTK} if we show that
 \begin{equation}
 \label{cn-mono}
 \int x^kd\bar{\nu}_{L,p}(x)\xrightarrow{L\to\infty} \int x^kd\bar{\nu}_p(x),~~\forall~k\in\mathbb{N}.
 \end{equation}
From the definition (\ref{inter}) of $\Lambda_{L,k}^{int}$, the interior of $\Lambda_L$, it is easy to see that for all 
$n\in\Lambda_{L,k}^{int}$ we always have
\begin{equation}
\label{eq-wh-fn}
\omega_n^{2p}\big\langle \delta_n, (H^\omega_L)^k\delta_n \big\rangle\big|_{(\omega_n\to u\omega_n)}=
\omega_n^{2p}\big\langle \delta_n, (H^\omega)^k\delta_n \big\rangle\big|_{(\omega_n\to u\omega_n)}.
\end{equation} 
Now using (\ref{mnt-thr-spec}) and the above we write
\begin{align}
\label{cp-cng}
\int x^kd\bar{\nu}_{L,p}(x)&=\frac{1}{|\Lambda_L|}\sum_{n\in\Lambda_L}\int_0^1\big[\mathbb{E}\big(\omega_n^{2p}\big\langle \delta_n, (H^\omega_L)^k\delta_n\big\rangle\big|_{(\omega_n\to u\omega_n)}\big)\big]du\nonumber\\
&=\frac{1}{|\Lambda_L|}\sum_{n\in\Lambda_{L,k}^{int}}\int_0^1\big[\mathbb{E}\big(\omega_n^{2p}\big\langle \delta_n, (H^\omega)^k\delta_n\big\rangle\big|_{(\omega_n\to u\omega_n)}\big)\big]du\nonumber\\
&\qquad ~+\frac{1}{|\Lambda_L|}\sum_{n\in \Lambda_L\setminus\Lambda_{L,k}^{int}}\int_0^1\big[\mathbb{E}\big(\omega_n^{2p}\big\langle \delta_n, (H^\omega_L)^k\delta_n\big\rangle\big|_{(\omega_n\to u\omega_n)}\big)\big]du\nonumber\\
&=\frac{1}{|\Lambda_L|}\sum_{n\in\Lambda_{L}}\int_0^1\big[\mathbb{E}\big(\omega_n^{2p}\big\langle \delta_n, (H^\omega)^k\delta_n\big\rangle\big|_{(\omega_n\to u\omega_n)}\big)\big]du\nonumber\\
&\qquad ~+\frac{1}{|\Lambda_L|}\sum_{n\in \Lambda_L\setminus\Lambda_{L,k}^{int}}\bigg(\int_0^1\big[\mathbb{E}\big(\omega_n^{2p}\big\langle \delta_n, (H^\omega_L)^k\delta_n\big\rangle\big|_{(\omega_n\to u\omega_n)}\big)\big]du\nonumber\\
&\qquad\qquad \qquad\qquad-\int_0^1\big[\mathbb{E}\big(\omega_n^{2p}\big\langle \delta_n, (H^\omega)^k\delta_n\big\rangle\big|_{(\omega_n\to u\omega_n)}\big)\big]du\bigg)\nonumber\\
&=\int_0^1\big[\mathbb{E}\big(\omega_0^{2p}\big\langle \delta_0, (H^\omega)^k\delta_0\big\rangle\big|_{(\omega_0\to u\omega_0)}\big)\big]du\nonumber\\
&\qquad~+\frac{1}{|\Lambda_L|}\sum_{n\in \Lambda_L\setminus\Lambda_{L,k}^{int}}\bigg(\int_0^1\big[\mathbb{E}\big(\omega_n^{2p}\big\langle \delta_n, (H^\omega_L)^k\delta_n\big\rangle\big|_{(\omega_n\to u\omega_n)}\big)\big]du\nonumber\\
&\qquad\qquad \qquad\qquad-\int_0^1\big[\mathbb{E}\big(\omega_n^{2p}\big\langle \delta_n, (H^\omega)^k\delta_n\big\rangle\big|_{(\omega_n\to u\omega_n)}\big)\big]du\bigg)\nonumber\\
&=\int_0^1\big[\mathbb{E}\big(\omega_0^{2p}\big\langle \delta_0, (H^\omega)^k\delta_0\big\rangle\big|_{(\omega_0\to u\omega_0)}\big)\big]du+\mathcal{E}_{k,L}\nonumber\\
&=\int x^kd\bar{\nu}_p(x)+\mathcal{E}_{k,L}.
\end{align}
 In the first part of the third line (from below) of the above, we have used the fact that the sequence of random variables
 $\bigg\{\omega_n^{2p}\big\langle \delta_n, (H^\omega)^k\delta_n\big\rangle\big|_{(\omega_n\to u\omega_n)}\bigg\}_{n\in\mathbb{Z}^d}$ has same distribution \big(one way to realise it,  consider $u$ is uniformly distributed on $[0,1]$ and independent of 
 $\{\omega_n\}_{n\in\mathbb{Z}^d}$\big).  Now using (\ref{est-ldsm}),  (\ref{bd-mnt-dm}) and (\ref{mnt-dm}) 
 the error term $\mathcal{E}_{k,L}$ can be estimated as
 \begin{equation}
 \label{err-st-mnt}
 \big| \mathcal{E}_{k,L} \big|\leq 2 (2d+1)^k C^k a^{k+2p} (k+2p)^{k+2p}~\frac{|\Lambda_L\setminus\Lambda_{L,k}^{int}|}{|\Lambda_L|}=O\big((2L+1)^{-1}  \big).
 \end{equation}
 Using (\ref{err-st-mnt}) in (\ref{cp-cng}) will give
 \begin{equation*}
 \int x^kd\bar{\nu}_L(x)\xrightarrow{L\to\infty}\int x^kd\bar{\nu}_p(x).
 \end{equation*}
 So we got (\ref{cn-mono}), hence the lemma.
\end{proof}
\noindent Now we define a sequence of random measure $\big\{\nu^\omega_L(\cdot)  \big\}_L$ as
\begin{equation}
\label{ldosmd}
\nu^\omega_L(\cdot)=\frac{1}{|\Lambda_L|}\sum_{n\in\Lambda_L}\big\langle \delta_n, E_{H^\omega_L}(\cdot)\delta_n\big\rangle.
\end{equation}
\noindent As we have seen in the Proposition \ref{mnt-det} under the condition (\ref{mnts}) on the moments of single site distribution  (SSD) $\mu(\cdot)$, the density of states measure (DOSm) $\nu(\cdot):=\mathbb{E}\big(\big\langle \delta_0, E_{H^\omega}(\cdot)\delta_0\big\rangle \big)$ is also will be determined by its moments (similar calculation),  see Corollary \ref{mnt-det-dosm}. Therefore, because of the result \cite[Theorem 30.2]{PB}, we can also talk about the weak convergence of the sequence of random probability measures $\{\nu^\omega_L(\cdot)\}_L$, defined by (\ref{ldosmd}).  
\begin{lem}
\label{poly-gr-weak}
Let the single site distribution (SSD) $\mu$ satisfy the moments condition (\ref{mnts}) and $f$ is a continuous function on $\mathbb{R}$ with $\big|f(x) \big|\leq P(x)~\forall~x\in\mathbb{R}$ for some polynomial $P$.  Then, there exists a set 
$\tilde{\Omega}\subset \Omega$ which is independent of $f$ and $\mathbb{P}(\tilde{\Omega})=1$ such that
\begin{equation}
\label{as-cng-poly-gr}
\frac{1}{|\Lambda_L|}\sum_{n\in\Lambda_L}\big \langle \delta_n,  f\big(H^\omega_L\big)\delta_n\big\rangle \xrightarrow {L\to\infty}
\mathbb{E}\big(\big \langle \delta_0,  f\big(H^\omega\big)\delta_0\big\rangle\big)~~\forall~\omega\in\tilde{\Omega}.
\end{equation}
\end{lem}
\begin{proof}
We use the definition of $\nu^\omega_L(\cdot)$ as in (\ref{ldosmd}) and the density of states measure (DOSm)
$\nu(\cdot):=\mathbb{E}\big(\big\langle \delta_0, E_{H^\omega}(\cdot)\delta_0\big\rangle \big)$ to write the above (\ref{as-cng-poly-gr}) as the convergence of integrals
\begin{equation}
\label{mes-th-cn}
\int f(x)d\nu^\omega_L\xrightarrow{L\to\infty} \int f(x)d\nu(x)~~\forall~\omega\in\tilde{\Omega}~~\text{with}~~\mathbb{P}\big(\tilde{\Omega}\big)=1.
\end{equation}
Since $f$ is a continuous function of polynomial growth and under the condition (\ref{mnts}), the density of state measure (DOSm) $\nu(\cdot)$ is determined by its moments (see Corollary \ref{mnt-det-dosm}), therefore, in the presence 
of \cite[Lemma 2.2 \& Lemma 2.1]{DTK} to prove the above (\ref{mes-th-cn}) it is enough to show 
\begin{equation}
\label{mnt-cn-equi}
\int x^kd\nu^\omega_L\xrightarrow{L\to\infty} \int x^k d\nu(x)~~\forall~k\in\mathbb{N}~~\text{and}~~\omega\in \tilde{\Omega}~~
\text{with}~~\mathbb{P}\big(\tilde{\Omega}  \big)=1,
\end{equation}
here $\tilde{\Omega}\subset \Omega$ and it is independent of $k$.\\
Now using (\ref{eq-wh-fn}) we write the $k$th moment $\nu^\omega_L(\cdot)$ as
\begin{align}
\label{exp-r-m}
\int x^kd\nu^\omega_L(x)&=\frac{1}{|\Lambda_L|}\sum_{n\in\Lambda_L}\big\langle \delta_n,  \big( H^\omega_L \big)^k\delta_n\big\rangle\nonumber\\
&=\frac{1}{|\Lambda_L|}\sum_{n\in\Lambda_{L,k}^{int}}\big\langle \delta_n,  \big( H^\omega \big)^k\delta_n\big\rangle\nonumber\\
&\qquad\qquad +\frac{1}{|\Lambda_L|}\sum_{n\in\Lambda_L\setminus\Lambda_{L,k}^{int}}\big\langle \delta_n,  \big( H^\omega_L \big)^k\delta_n\big\rangle\nonumber\\
&=\frac{1}{|\Lambda_L|}\sum_{n\in\Lambda_L}\big\langle \delta_n,  \big( H^\omega \big)^k\delta_n\big\rangle\nonumber\\
&\qquad\qquad +\frac{1}{|\Lambda_L|}\sum_{n\in\Lambda_L\setminus\Lambda_{L,k}^{int}}\bigg(\big\langle \delta_n,  \big( H^\omega_L \big)^k\delta_n\big\rangle
-\big\langle \delta_n,  \big( H^\omega \big)^k\delta_n\big\rangle\bigg)\nonumber\\
&=\frac{1}{|\Lambda_L|}\sum_{n\in\Lambda_L}\big\langle \delta_n,  \big( H^\omega \big)^k\delta_n\big\rangle+\mathcal{E}_{L,k}(\omega).
\end{align}
Since the coefficients $C^{j_1,j_2,\cdots, j_k}_{n_1,n_2,\cdots n_k, L}$ in (\ref{mono-fn}), the expression of $\big\langle \delta_n, \big( H^\omega_L \big)^k \delta_n\big\rangle$ are smaller the coefficients $C^{j_1,j_2,\cdots, j_k}_{n_1,n_2,\cdots n_k}$ in (\ref{mono}),  the expression of $\big\langle \delta_n, \big( H^\omega \big)^k \delta_n\big\rangle$ and the moments of $\mu(\cdot)$ (DOSm) is bounded by (\ref{mnts}).  Therefore it is possible to find a constant $M_k>0$ (independent of $L$) such that
\begin{equation}
\label{var-fn-est}
\mathbb{E}\big(\big( \mathcal{E}_{k,L} \big) ^2 \big)\leq M_k~\bigg(\frac{|\Lambda_L\setminus\Lambda_{L,k}^{int}|}{|\Lambda_L|}\bigg)^2
=O\big((2L+1)^{-2}  \big).
\end{equation}
 The use of Chebyshev's inequality and the above will give
\begin{equation}
\label{sum-fn-bo}
\sum_{L\ge 1}\mathbb{P}\big(\omega:   \big|\mathcal{E}_{k,L}(\omega)\big|> \epsilon\big)<\infty,~~\epsilon>0~~\text{for each}~k\in\mathbb{N}.
\end{equation}
Now the Borel$-$Cantelli lemma will ensure the almost sure convergence of the sequence of random variables $\{\mathcal{E}_{k,L}\}_L$ to the zero, i.e
\begin{equation}
\label{as-0}
\mathcal{E}_{k,L}(\omega)\xrightarrow {L\to\infty} 0~~a.e~\omega.
\end{equation}
It is also true that $\big\{\big\langle \delta_n,  \big( H^\omega\big)^k\delta_n \big\rangle \big\}_{n\in\mathbb{Z}^d}$ is an ergodic process therefore 
\begin{equation}
\label{er-cng}
\frac{1}{|\Lambda_L|}\sum_{n\in\Lambda_L}\big\langle \delta_n,  \big( H^\omega\big)^k\delta_n \big\rangle\xrightarrow {L\to\infty}
\mathbb{E}\big(\big\langle \delta_0,  \big( H^\omega\big)^k\delta_0 \big\rangle  \big)~~a.e~\omega.
\end{equation}
Use of (\ref{er-cng}), (\ref{as-0}) and the definition of $\nu(\cdot)$ (DOSm) in (\ref{exp-r-m}) will give
\begin{equation}
\label{as-mnt-cng}
\int x^kd\nu^\omega_L(x)\xrightarrow {L\to\infty}\int x^k d\nu(x)~~\forall~\omega\in \Omega_k~~\text{with}~~\mathbb{P}\big(\Omega_k  \big)=1.
\end{equation}
Define $\tilde{\Omega}=\displaystyle\bigcap_{k\in\mathbb{N}} \Omega_k$ then (\ref{mnt-cn-equi}) is immediate from the above.  Hence, the proof of the lemma is done.\end{proof}
\begin{rem}
As we have seen in the Lemma \ref{poly-gr-weak} that under the moment condition (\ref{mnts}) on the single site distribution (SSD) $\mu(\cdot)$ the convergence (\ref{ids}) will hold for a much larger class, namely set of all continuous function with polynomial growth.  
\end{rem}

\noindent {\bf Acknowledgements}: The author is partially supported by the INSPIRE faculty fellowship grant of the Department of Science and Technology,  Government of India. The author is thankful to M. Krishna for his valuable comments.

\end{document}